\documentclass{article}
\usepackage[utf8]{inputenc}
\usepackage{amsmath,esint,mathtools}
\usepackage{amssymb}
\usepackage[arrowdel]{physics}
\usepackage{bm}
\usepackage{savesym}
\savesymbol{checkmark}
\usepackage{dingbat}
\usepackage{graphicx}
\usepackage{caption}
\usepackage[bookmarks=false]{hyperref}
\hypersetup{colorlinks,allcolors=black}
\usepackage{geometry}
\usepackage{sectsty}
\newcommand{\uvec}[1]{\boldsymbol{\hat{{#1}}}}
\makeatletter
\newcommand{\setword}[2]{%
  \phantomsection
  #1\def\@currentlabel{\unexpanded{#1}}\label{#2}%
}
\makeatother

\NewDocumentCommand\interior{sm}
  {\IfBooleanTF{#1}{?}{\mathring{#2}}{}}
\title{Circuit quantization with time-dependent flux: the parallel-plate SQUID}
\author{$^1$ Rohan Narayan Rajmohan, $^2$  Ahmed Kenawy \& $^2$ David DiVincenzo \and $^1$ Indian Institute of Technology Madras, Chennai-600036, India \and $^2$ Peter Grünberg Institute (PGI-2), Forschungszentrum Jülich, D-52425, Jülich, Germany}

\date{}
\begin{document}
\maketitle
\begin{abstract}
Quantum circuit theory has emerged as an essential tool for the study of the dynamics of superconducting circuits. Recently, the problem of accounting for time-dependent driving via external magnetic fields was addressed by Riwar-DiVincenzo in their paper\textemdash`Circuit quantization with time-dependent magnetic fields for realistic geometries' in which they proposed a technique to construct a low-energy Hamiltonian for a given circuit geometry, taking as input the external magnetic field interacting with the geometry. This result generalises previous efforts that dealt only with discrete circuits. Moreover, it shows through the example of a parallel-plate SQUID circuit that assigning individual, discrete capacitances to each individual Josephson junction, as proposed by treatments of discrete circuits, is only possible if we allow for negative, time-dependent and even singular capacitances. In this report, we provide numerical evidence to substantiate this result by performing finite-difference simulations on a parallel-plate SQUID. We furnish continuous geometries with a uniform magnetic field whose distribution we vary such that the capacitances that are to be assigned to each Josephson junction must be negative and even singular. Thus, the necessity for time-dependent capacitances for appropriate quantization emerges naturally when we allow the distribution of the magnetic field to change with time.
\end{abstract}

\section{Introduction}
The paper by You et al. addressed the issue that the important aspects of the Hamiltonian description of cQED, when accounting for time-dependent fluxes, have not been appropriately discussed. It brought up the fact that the system dynamics differ based on the choice of basis due to the presence of a term linearly dependent on the time derivative of the total external flux that threads the circuit \textemdash $\dot{\Phi}_\text{e}$. For special choices of basis, this term vanishes and all that remains is a parametric dependence on time where the time-independent Hamiltonian H($\phi$) is modified to H($\phi$(t)). In Ref.~\hyperref[sec:Ref]{2}, this basis was found for lumped-element quantum circuits by means of an `irrotational' gauge for the Hamiltonian. 
\vspace{2mm}
\\
The paper by Riwar-DiVincenzo generalized the notion of an irrotational gauge to continuous circuit models and found that the suitable generalization is the Coulomb gauge, supported by additional boundary conditions at the superconducting surfaces \textemdash in accordance with the London gauge constraints. The paper then established the connections between this vector potential gauge for continuous geometries and the irrotational gauge developed in Ref.~\hyperref[sec:Ref]{2} for discrete circuits. Ref.~\hyperref[sec:Ref]{1} established that even for a simple SQUID geometry, assigning individual capacitances to each Josephson junction for the modelling of the continuous geometry as a lumped-element system leads to a requirement for negative, time-dependent and even momentarily singular capacitances. 
\vspace{2mm}
\\
This work, through numerical simulations, reinforces the findings of Ref.~\hyperref[sec:Ref]{1} and proves that continuous geometries do \textbf{not} generalize easily to the predictions of Ref.~\hyperref[sec:Ref]{2} and that the capacitances that we must assign to each junction are, in some cases, not equal to the electrostatic junction capacitances associated with the structure and must depend on the spatial position of the magnetic field.

        \begin{figure}[h]
            \centering
            \includegraphics[width=8 cm]{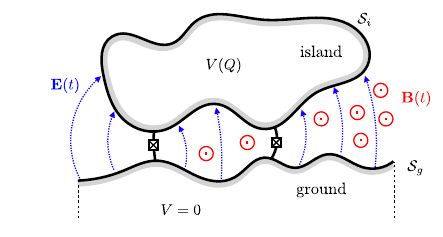}
            \qquad
            \caption{\small The general geometry of the circuits considered - with the time-dependent flux, B(t), as input in general, from which the time-dependent electric field is derived. We are only concerned with single snapshots of the evolution and thus do not consider the time-dependence of the field nor the associated electric field.}
            \label{fig:1}
        \end{figure}
\newpage
\maketitle
\section{Gauge fixing for a general geometry}
\begin{figure}[h]
   \centering
   \includegraphics[width=8 cm]{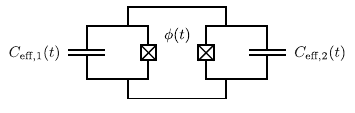}
    \qquad
    \caption{\small It is found that in general, the continuous geometry of Fig.~\hyperref[fig:1]{1} cannot be reduced to this lumped-element equivalent unless we allow for time-dependent, negative and potentially singular capacitances for $C_\text{eff,1}(t)$ and $C_\text{eff,2}(t)$}
    \label{fig:2}
\end{figure}

\vspace{2 mm}
Fig.~\hyperref[fig:1]{1} represents a general structure of a SQUID circuit, driven by a time-dependent magnetic field \textbf{(in red)}, which creates the electric field \textbf{(in blue)}. The lumps are perfect superconductors, expelling both electric and magnetic fields from their interior, and are referred to as 'the island' and 'the ground' respectively. The voltage of the ground is set to 0. They have surfaces, $S_l$, with vector $\uvec{n}_l$ normal to the surface. The index \textit{l=i,g} enumerates the two superconductors. The two Josephson junctions that connect the island and the ground enclose a finite area, A. 
\vspace{2mm}
\\
The flux through this area is given by \textemdash
\begin{equation}
    \Phi(t)=  \iint\limits_A \uvec{n}_A \cdot \Vec{B}(\Vec{r},t)\,dA \hspace{2 mm}. 
    \label{eqn:1}
\end{equation}
\vspace{2 mm}
\\
Circuits such as those in Fig.~\hyperref[fig:1]{1} can be transformed to their lumped-element equivalent in Fig.~\hyperref[fig:2]{2} where to each Josephson junction with an energy $E_{J_k}$, we assign a capacitance $C_k$.The external flux, $\Phi$(t) enters the Hamiltonian as a phase \textemdash $\phi$(t)=$2\pi\Phi(t)/ \Phi_0$, with $\Phi_0$ being the flux quantum \textemdash  $h/(2e)$
\vspace{2mm}
\\
For a time-independent flux, $\dot{\phi}=0$, the dynamics of the SQUID circuit is described by a Hamiltonian :- 
\begin{equation}
    \uvec{H}_{\alpha}=\frac{(2e)^2}{2C_{tot}}.\uvec{n}^2-E_{J_1} cos(\uvec{\Phi}-\alpha*\phi)-E_{J_2} cos(\uvec{\Phi}-(\alpha-1)*\phi) \hspace{2 mm}.
    \label{eqn:2}
\end{equation}
\vspace{2mm}
\\
In the situation where $\dot{\phi}=0$, the real parameter $\alpha$ can be chosen arbitrarily and the different Hamiltonians \textemdash {$\uvec{H}_{\alpha}$} are connected via unitary transformations : $\uvec{H}_{\alpha}=\uvec{U}_{\alpha}\uvec{H}_{0}\uvec{U}^\dagger_{\alpha}$ : $\uvec{U}_\alpha=e^{i \alpha \phi \uvec{n}}$. This is a gauge freedom, equivalent to the gauges in continuous-domain electromagnetism. However, if there is a time-dependent driving field - $\Vec{B}(t)$, resulting in a time-dependent flux ($\dot{\phi}$), for a general $\alpha$, it is \textbf{not} possible to simply replace $\phi$ with $\phi$(t) in the Hamiltonian Eq.~(\hyperref[eqn:2]{2}) and arrive at the Hamiltonian for the time-dependent flux case. Thus, Hamiltonians Eq.~(\hyperref[eqn:2]{2}) with different $\alpha$ give rise to physically different time evolution - which arises due to the fact that the unitary transformations that connect these hamiltonians is now time-dependent, and as expressed in both Ref.~\hyperref[sec:Ref]{1} and Ref.~\hyperref[sec:Ref]{2}, appears in the Schrödinger equation as the extra term - $-i \uvec{U}_\alpha \partial_t \uvec{U}^\dagger_\alpha  = -\alpha \dot{\phi} \uvec{n}$. Hence, to accurately capture the dynamics of the time-dependent problem, $\alpha$ must be fixed. 
\vspace{2mm}
\\
It is shown in Ref.~\hyperref[sec:Ref]{2} that for a SQUID, setting $\phi$ $\rightarrow \phi$(t) in Eq.~(\hyperref[eqn:2]{2}) accurately captures the dynamics of the time-dependence if :\textemdash
\begin{equation}
    \alpha=\frac{C_2}{C_\text{tot}}
    \label{eqn:3}
\end{equation}
\vspace{2mm}
This is referred to as `the irrotational gauge' by the authors of Ref.~\hyperref[sec:Ref]{2}. For other gauges, terms linear in $\dot{\phi}$ appear. 

\maketitle
\section{Aim}
In Ref.~\hyperref[sec:Ref]{1}, the assumption that it is possible to assign definite, positive capacitances to each junction, while transitioning from the continuous geometry to the lumped-element equivalent, that only depend on the geometry of the structure is questioned and it is proven in Ref.~\hyperref[sec:Ref]{1} that this hypothesis \textbf{must} be relaxed as the mapping from Fig.~\hyperref[fig:1]{1} to Fig.~\hyperref[fig:2]{2} is only possible if one allows for potentially negative, time-dependent and even \textbf{momentarily singular} capacitance values for $C_\text{eff,1}$(t) \& $C_\text{eff,2}$(t) that depend on the continuous geometry of the structure as well as the spatial distribution of the magnetic field with the only guarantee being that $C_\text{tot}=\sum_k C_\text{eff,k}$(t) is constant and positive, depending only on the geometry. $C_\text{tot}$ being positive \textbf{and} constant still allows a parametric variation of $C_\text{eff,1} \& C_\text{eff,2}$ with time, which is sufficient to ensure that the `irrotational condition' is satisfied.
\vspace{2mm}
\\
The purpose of this work is to reaffirm the conclusions of Ref.~\hyperref[sec:Ref]{1} via numerical simulations for a continuous geometry whose cross-section in the x-y plane is shown in Fig.~\hyperref[fig:4]{4} - extended in the same manner, uniformly for $z=\pm \infty$. We vary the spatial distribution of a uniform magnetic field oriented perpendicular to the x-y plane, to examine variation in $\alpha$.
\vspace{2mm}
\\
Since the structure extends in a symmetric manner along the z-axis, the analysis we have to perform needs only be done in 2-D as any `slice' of the structure will behave in the same manner due to translational symmetry along the z-axis.
\maketitle
\section{Methodology}
\label{sec:4}
Since the system only has a parametric dependence on time, it suffices to do a time-independent analysis and perform a comparative analysis of the $\alpha$ values.The numerical simulations are performed using Mathematica, using a finite difference approach- approximating the derivatives using the central derivative approach in standard finite difference calculations. It is shown in Sec.~\hyperref[sec:5]{5} \& Sec.~\hyperref[sec:6]{6} that the equations and the boundary conditions involved only require derivatives up to the second order. Thus we only require finite approximations to derivatives up to the second order. Mixed derivatives are not required.
\vspace{2 mm}
\\
\textbf{Finite approximations to the derivatives :-}
\\
$h_x$ - Size of a division in the x-dimension
\\
$h_y$ - Size of a division in the y-dimension
\vspace{2mm}
\\
\begin{align*}
\frac{\partial (\psi(x,y))}{\partial x}|_{(x_0,y_0)} & \rightarrow \frac{\psi(x_0+h_x,y_0)-\psi(x_0-h_x,y_0)}{2*h_x}\\
\frac{\partial (\psi(x,y))}{\partial y}|_{(x_0,y_0)}  & \rightarrow \frac{\psi(x_0,y_0+h_y)-\psi(x_0,y_0-h_y)}{2*h_y}\\
\frac{\partial^2 (\psi(x,y))}{\partial x^2}|_{(x_0,y_0)} & \rightarrow \frac{\psi(x_0+h_x,y_0)+\psi(x_0-h_x,y_0)-2*\psi(x_0,y_0)}{h_x^2}\\
\frac{\partial^2 (\psi(x,y))}{\partial y^2}|_{(x_0,y_0)} & \rightarrow \frac{\psi(x_0,y_0+h_y)+\psi(x_0,y_0-h_y)-2*\psi(x_0,y_0)}{h_y^2}\\
\end{align*}
Neumann boundary conditions are used for the superconducting surfaces.
\\
$\infty$ is treated by approximating it to a finite value, `far enough' from the system and implementing the fall-off conditions via a set of Dirichlet conditions on the finite boundary encapsulating the solution space.Checks were done to ensure that the finitude of the $\infty$ approximation does not force the solution to fall-off at a faster rate or behave incongruously.
\vspace{2 mm}
\\
\textbf{Interpolation Method :\textemdash}
The ListInterpolation[\hspace{1 mm}] function in Mathematica is used for interpolation to construct a function that takes on the values of the solution space at the grid points. In the tabulated results, all loop integrals for the total flux are calculated via discrete summation as an integral will trivially always yield the area enclosed itself. The $\alpha$ values are calculated via both the summation method and via an integral \textemdash to provide an estimate for errors. Values obtained by manipulating the interpolated solution are denoted by the tag \textbf{IM} and those obtained by discrete summations are denoted by the tag \textbf{SM}.

\maketitle
\section{Boundary conditions on the vector potentials}
\label{sec:5}
As stated in Sec.~\hyperref[sec:4]{4}, the numerical simulations performed are time-independent. Thus, the magnetic field is `static' in the sense that there is no associated electric field that needs to be considered. Thus, in terms of the vector potentials, we only have to deal with the magnetic vector potential.
\vspace{2 mm}
\\
The irrotational gauge, for continuous geometries, reduces to the Coulomb gauge according to Ref.~\hyperref[sec:Ref]{1} and is supplemented by appropriate boundary conditions at the superconducting surfaces.
\vspace{2 mm}
\\
As in Ref.~\hyperref[sec:Ref]{1}, $\lambda$ - the penetration depth - has been assumed to be negligible in comparison to the dimensions considered, and thus, the irrotational gauge loses its connection to the London gauge and the absolute value of the vector potential \textbf{within} the bulk of no concern with regard to the problem at hand - as long as magnetic field is expelled from the bulk of the superconductors in accordance with the Meissner effect.
\vspace{2mm}
\\
Thus, the vector potential $\Vec{A}_{irr}(x,y)$ must satisfy the following conditions :\textemdash

\begin{equation}
    \nabla \cross \Vec{A}(x,y) = 0 - \forall \text{(x,y) in the bulk of the superconductor}\hspace{2 mm},
\end{equation}
\begin{equation}
    \uvec{n}_l \cross \Vec{A}(x,y) = 0 - \forall \text{(x,y) $\in S_l$ - the set of all points on the surface of the superconductors}\hspace{2 mm}.
\end{equation}
\maketitle
\section{The stream function}
\label{sec:6}
\vspace{2 mm}
The Lagrange stream function, thereafter referred to as `the stream function'  without ambiguity, is a 2-D function that characterizes divergence-free flow in 2-D with the `flow velocity' components being expressed as derivatives of the stream function. It characterizes all the aspects of a divergence-free flow in 2-D and can be used to plot streamlines.
\vspace{2mm}
\\
The stream function is defined as :-
\begin{align}
  \Vec{A}_\text{irr} &=\uvec{z} \cross \nabla \psi \hspace{2 mm}, \\
  \implies \Vec{A}_\text{irr} &=\frac{\partial \psi}{\partial x}\uvec{y}-\frac{\partial \psi}{\partial y}\uvec{x}\hspace{2 mm}.
\end{align}
\newline
The divergence-free nature of the magnetic vector potential is implicit in the definition of $\Vec{A}_\text{irr}$ :\textemdash
\begin{equation*}
    \nabla \cdot \Vec{A}_\text{irr}=\frac{\partial}{\partial y}(\frac{\partial \psi}{\partial x})+ \frac{\partial}{\partial x}(-\frac{\partial \psi}{\partial y})=0 \hspace{2 mm}.
\end{equation*}
\newline
Then, the equation - $\nabla \cross \Vec{A}_\text{irr}=\Vec{B} = B(x,y) \uvec{z}$ is equivalent to the scalar equation:\textemdash
\begin{equation}
    \frac{\partial^2 \psi}{\partial x^2}+\frac{\partial^2 \psi}{\partial y^2}=B(x,y) \hspace{2 mm} \forall \text{(x,y) in the solution space}\hspace{2 mm}.
\end{equation}
The boundary condition at the surface of the superconductor reduces to:\textemdash
\begin{equation}
    \uvec{n}_l \cross (\uvec{z} \cross \nabla \psi) = 0 \hspace{2 mm},
\end{equation}
$\implies$
\begin{equation}
     \uvec{n}_l \cdot \nabla \psi= 0 \hspace{2 mm}.
\end{equation}

\maketitle
\section{A rejigged notation}
\vspace{2 mm}
In Ref.~\hyperref[sec:Ref]{[1]}, Eq.~(\hyperref[eqn:2]{2}) is rewritten as :-
\vspace{2 mm}
\begin{equation}
    \uvec{H}_{\alpha}=\frac{(2e)^2}{2C_\text{tot}}.\uvec{n}^2-E_{J_1} cos(\uvec{\Phi}+\phi_\text{1,irr})-E_{J_2} cos(\uvec{\Phi}+\phi_\text{2,irr}) \hspace{2 mm},
    \label{eqn:11}
\end{equation}
\\
with the phases that appear inside the potential energy terms of Eq.~(\hyperref[eqn:11]{11}), in the irrotational gauge at the junction k given through the line integral:\textemdash
\vspace{2 mm}
\begin{equation}
    \phi_\text{k,irr}=\frac{2\pi}{\Phi_0}\int_{L_{J_k}}\Vec{dl}\cdot\Vec{A}_\text{irr} \hspace{2 mm},
    \label{eqn:12}
\end{equation}
where $L_{J_\text{k}}$ is the shortest path across the junction k.
\newline
The irrotational vector potential not being curl-free by definition implies that the phases in Eq.~(\hyperref[eqn:11]{11}) depends on the path.
\medskip
Ref.~\hyperref[sec:Ref]{1} justifies the choice of this path as $L_{J_k}$ as :-
\medskip
\begin{enumerate}
    \item Large deviations from the shortest path are extremely unlikely due to standard path integral considerations.
    \item If fashioned for quantum hardware purposes, the sizes of Josephson junctions will be such that the magnetic flux penetrating them is much smaller than the flux quantum. Then, small deviations from the shortest path will leave the integral unchanged.
\end{enumerate}
Solving for theses phases numerically and then comparing Eq.~(\hyperref[eqn:11]{11}) to Eq.~(\hyperref[eqn:2]{2}) gives us the correct junction capacitances themselves.
\vspace{2 mm}
In this work, we define two parameters \textemdash $\alpha_1$ \& $\alpha_2$ \textemdash which serve as re-normalized versions of Eq.~(\hyperref[eqn:12]{12}) :\textemdash
\vspace{1mm}
\begin{equation}
    \alpha_{1}=\int_{L_{J_1}}\Vec{dl}\cdot\Vec{A}_\text{irr}\hspace{2 mm},
    \label{eqn:13}
\end{equation}
\vspace{1mm}
\begin{equation}
    \alpha_{2}=\int_{L_{J_2}}\Vec{dl}\cdot\Vec{A}_\text{irr}\hspace{2 mm}.
    \label{eqn:14}
\end{equation}
The integral in Eq.~(\hyperref[eqn:13]{13}) is defined from the island to the ground while Eq.~(\hyperref[eqn:14]{14}) is defined from the ground to the island. The directions are so defined so that on completing the loop within the superconductors, where $\Vec{A}=0$ as per the London equations, the loop integral will evaluate to the total flux enclosed by the Josephson junctions, $\Phi$.
\begin{equation}
    \therefore \hspace{8 mm} \frac{\alpha_1+\alpha_2}{\Phi}=1 \hspace{2 mm}.
    \label{eqn:15}
\end{equation}
This definition of $\alpha_1$ \& $\alpha_2$ results in these mappings to the notation used in Ref.~\hyperref[sec:Ref]{[2]} and Ref.~\hyperref[sec:Ref]{[1]}.
\begin{align}
\label{eqn:16}
    \alpha_1 &=-\phi_\text{1,irr} \frac{\Phi_0}{2\pi}=\alpha \Phi \hspace{2 mm},\\
\label{eqn:17}
    \alpha_2 &=+\phi_\text{2,irr} \frac{\Phi_0}{2\pi}=(1-\alpha) \Phi \hspace{2 mm}.
\end{align}
\vspace{2mm}
\\
In terms of the stream function, the $\alpha_1$ \& $\alpha_2$ integrals reduce to:-
\begin{align}
    \alpha_1 &=\int_{L_{J_k:i\rightarrow g}}\nabla\psi \cdot (\Vec{dl}\cross\uvec{z}) \hspace{2 mm}, \\
    \alpha_2 &=\int_{L_{J_k:g\rightarrow i}}\nabla\psi \cdot (\Vec{dl}\cross\uvec{z}) \hspace{2 mm}.
\end{align}
\maketitle
\section{Defining the geometry}
\begin{figure}[h]
    \centering
    \includegraphics[width=8 cm]{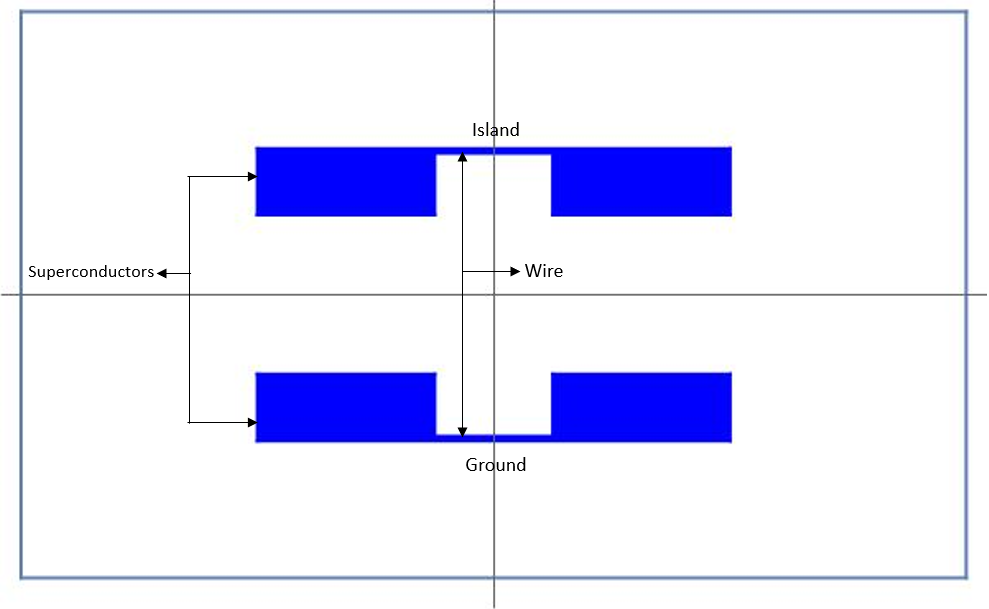}
    \qquad
    \caption{\small The cross-section of the parallel-plate SQUID structure. This cross-section extended uniformly from z=-$\infty$ to z=+$\infty$ forms the structure under consideration - although, in most of this article, we will deal with just the 2-D cross-section of the structure on the x-y plane and thus, for most practical purposes, abstract over the extension to z=$\pm \infty$. We take the assumption of a finite rest condition on the boundary shown, which is the finite approximation of `infinity'.   Throughout this article, we refer to the central gap in the structure as the `U-shaped' region. This region `splits' the structure into two `parallel-plate capacitors' - $C_1$, on the left, and $C_2$, on the right. To ensure strong equivalence with the parallel-plate setting, the dimensions of the `wire' are made extremely small. Note that this is just one of the dimensions of the parallel-plate SQUID that will be considered and is merely representative.}
    \label{fig:3}
\end{figure}

The structure consists of two `U-shaped' superconductors, as shown in Fig.~\hyperref[fig:3]{3}, which is representative of two parallel plate capacitors connected by a thin wire. The Josephson junctions are located in the gap between the plates of each capacitor and define the position of the line integrals for $\alpha_1,\alpha_2$ by altering the two integrals when the magnetic field is placed in the gap between the plates of a capacitors.
\vspace{2mm}
\\
The requirement that the line integral of the vector potential over any closed loop be equal to the flux enclosed and the fact that the magnetic field in our problem is localized within a finite region implies that the vector potential must asymptotically tend to $\Vec{0}$. However, the finitude of the solution space implies that the behaviour of the stream function at infinity must be enforced on a finite boundary. Thus, the vector potential must vanish on the boundary of the solution space. This corresponds to providing Dirichlet conditions on the bounding curve of the solution space. The vector potential vanishing on the boundary implies that the stream function must have vanishing derivatives on the boundary. Thus, the stream function takes a constant value over this boundary, which we take to be zero, without any loss of generality. Assigning any other value to the stream function on the boundary will only result in a linear shift of the entire solution.
\vspace{2mm}
\\
\textbf{Note:\textemdash }To ensure that the structure in Fig.~\hyperref[fig:3]{3} is representative of two parallel-plate capacitors connected by a thin wire, the thickness of the `connecting wire' must be significantly less than the dimensions of the capacitive plates.
\vspace{2mm}
\\
Since Ref.~\hyperref[sec:Ref]{1} expressly deals with `physically realizable' settings, the field we consider will be a constant, uniform magnetic field constrained within a rectangular region. This type of field can be associated with that produced by a current carrying solenoid.
\maketitle
\section{Problem definition}
\vspace{2 mm}
\subsection{Parameters for grid definition}
        \begin{enumerate}
            \item $\{x_{min},x_{max}\}$ - The extents of the x-axis
            \item $\{y_{min},y_{max}\}$ - The extents of the y-axis
            \item $\{ \text{xdivisions,ydivisions} \}$  - Controls the grid resolution
            \item $\{ h_x,h_y \}$ - The $\Delta$\textit{x} $\& \Delta$\textit{y} equivalents
        \end{enumerate}
\subsection{Grid specification}
For the numerical simulations performed, the grid is chosen such that:-
\begin{enumerate}
    \item The boundary is at a sufficiently large enough distance from the superconductors to ensure that the approximation of `infinity' as a finite boundary does not affect the solution beyond an acceptable tolerance limit. 
    \item The grid must be fine enough to capture the entire behaviour of the solution and ensure that using a finer grid does not alter the solution significantly.
\end{enumerate}
\newpage
Thus, we fix the grid parameters as:-
        \begin{enumerate}
            \item $\{x_\text{min},x_\text{max}\} = \{-1,1\}$
            \item $\{y_\text{min},y_\text{max}\} = \{-1,1\}$
            \item $\{ \text{xdivisions,ydivisions} \} = \{128,128\}$
            \item $\{ h_x,h_y \} = \{\frac{1}{64},\frac{1}{64}\}$
        \end{enumerate}
\vspace{2 mm}
\subsection{Parameters for structural definition}
\begin{figure}[h]
    \centering
     \includegraphics[width=8 cm]{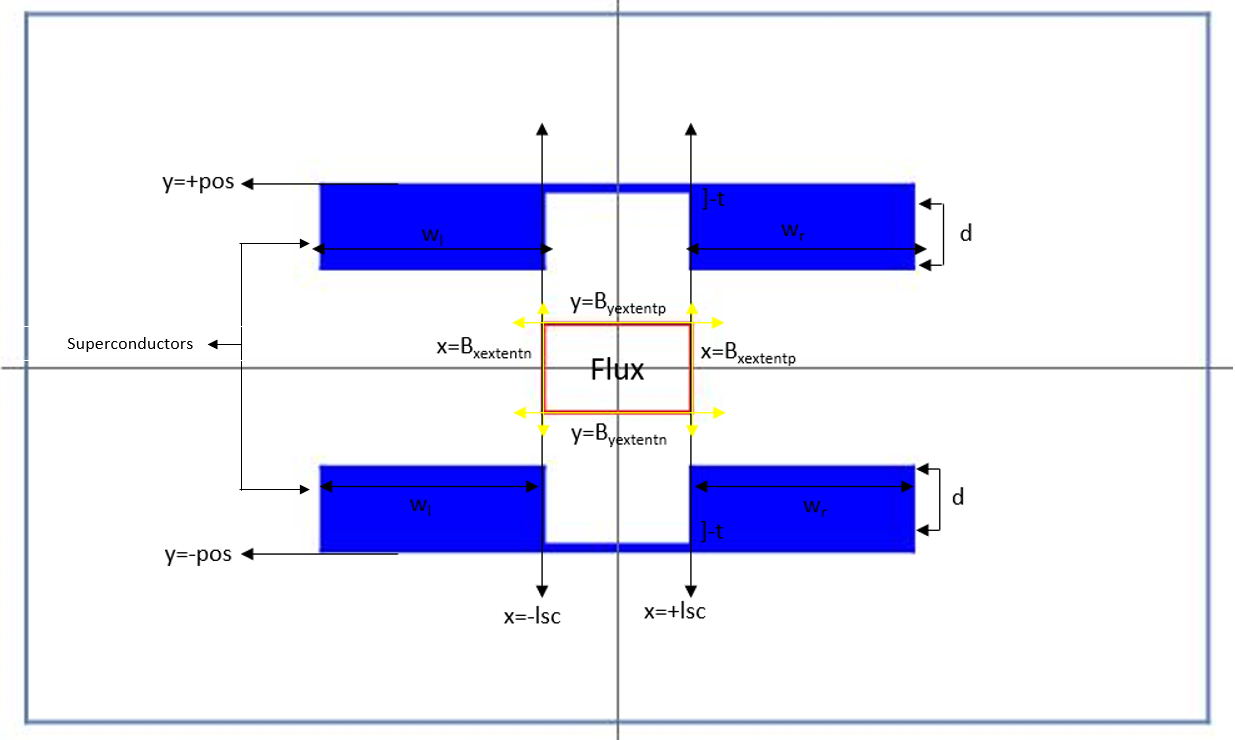}
    \qquad
    \caption{\small The parameters and the parts of the continuous geometry they control.}
    \label{fig:4}
\end{figure}

Fig.~\hyperref[fig:4]{4}, maps the parameters that specify the SQUID structure to the parts of the fixture they control. The flux is concentrated over a rectangular region, specified by 4 parameters.
\vspace{1mm}
\begin{enumerate}
    \item $\{ \text{B}_\text{xextentn},\text{B}_\text{xextentp} \}$ $\rightarrow$ The x-extents of the magnetic field region
    \item $\{ \text{B}_\text{yextentn},\text{B}_\text{yextentp} \}$ $\rightarrow$ The y-extents of the magnetic field region
    \item lsc $\rightarrow$ The length of the `wire' in the x-direction, symmetric about $y=0$
    \item $w_\text{l}$ $\rightarrow$ The width of the `capacitor' on the left
    \item $w_\text{r}$ $\rightarrow$ The width of the `capacitor' on the right
    \item d $\rightarrow$ The length of the `capacitances' in the y-dimension
    \item pos $\rightarrow$ The y-coordinate of the `bottom surface' of the wire
    \item t $\rightarrow$ The thickness of the `wire'
\end{enumerate}
\vspace{2mm}

\maketitle
\section{Computation of the electrostatic junction capacitances}
To compare the $\alpha$ values obtained numerically to the predictions of Ref.[2], it is necessary to evaluate the electrostatic junction capacitances. In the setting of our problem, it is possible to delimit the capacitances and two natural choices arise for the same. The first choice is to choose $x=\pm lsc$ as the delimiters. This corresponds to our view of the structure being the rough equivalent of two parallel plate capacitors connected via a thin wire. The alternative is to assign the role of the delimiter to the surface $x=0$. This approach is prudent as it then leads to a straightforward definition of $C_\text{tot}$ and accounts for charge formation on the wire and thus, the entire structure. 
\vspace{2mm}
\\
The second approach does not lead to a marked deviation from the first as the charge on the wire is negligible in comparison to that on the `plates'.This ensures that the theoretical equivalence to the `two plates joined by a wire' is still maintained. Thus, in this paper, we delimit the electrostatic junction capacitances at the surface $y=0$. 
\vspace{2mm}
\\
To stress on the approximate equivalence of the two methods of delimiting the junction capacitances, in Sec.~\hyperref[sec:12]{12} and Sec.~\hyperref[sec:13.2.1]{13.2.1}, electrostatic junction capacitances obtained by the two methods are contrasted.
\vspace{2mm}
\\
The capacitances are computed via the standard numerical technique of maintaining the two surfaces at constant potential and evaluating the Gaussian surface integrals.
\vspace{2mm}

\maketitle
\section{Tests for consistency}
\subsection{Loop integral tests}
\vspace{2mm}
Consider a closed surface, S, enclosed by a loop L : L=$\partial$S. Then, by the Gauss theorem - $\oint_{L} \Vec{A}_\text{irr}\cdot \Vec{dl}= \iint_{S} B$ \hspace{1 mm} dS \hspace{2 mm}.
\vspace{2mm}
\\
In terms of the stream function, this is equivalent to:\textemdash
\begin{equation}
    \oint_{L} \nabla \psi \cdot (\Vec{dl}\cross\uvec{z})=\iint_{S} B \hspace{1 mm} dS \hspace{2 mm}.
\end{equation}
\vspace{2 mm}
\\
This property must be satisfied, within acceptable tolerance limits, by any loop that lies within the solution space. Using this, we can generate a series of self-consistency checks. 
\vspace{2mm}
\\
We choose the following 3 loops and evaluate the line integral on the closed loops \textemdash both as a finite summation of the discrete solution yielded by the finite difference method and as an integral of the interpolated, continuous solution.
    \begin{enumerate}
        \item Loop 1 : x=$\frac{1}{2} : y \in [-1,1]  (-\uvec{y}) \cup x=          \frac{1}{2} : y \in [-1,1] (+\uvec{y}) \cup y=-1 : x \in                [\frac{-1}{2},\frac{1}{2}] (+\uvec{x}) \cup y=1 : x \in                 [\frac{-1}{2},\frac{1}{2}] (-\uvec{x})$
            \begin{figure}[h]
              \centering
              \includegraphics[width=8 cm]{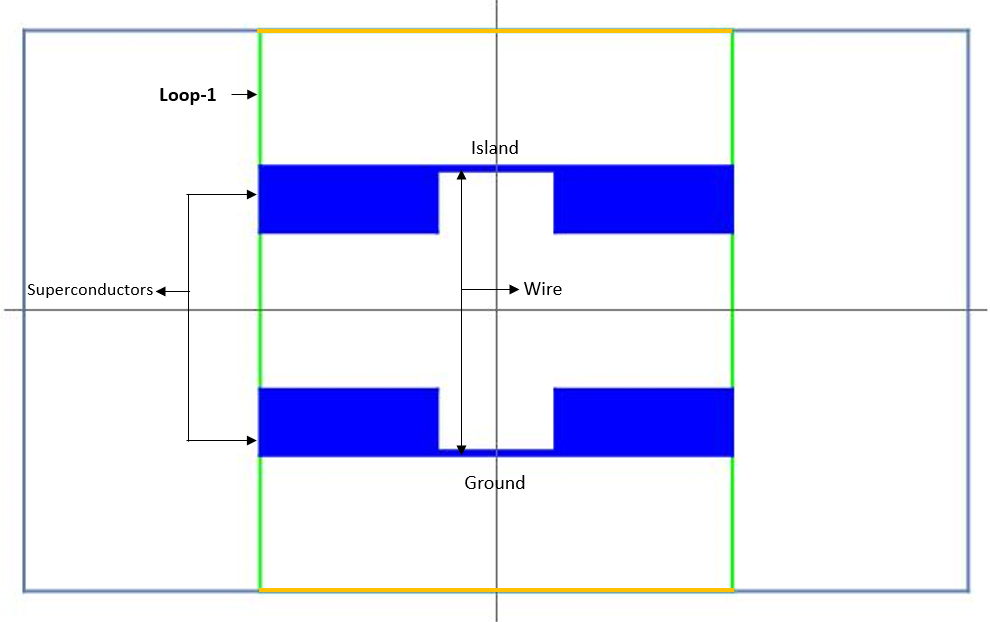}
              \qquad
              \caption{\small $L_1$ is completed at the finite boundary we approximate to be `infinity'. Thus, derivatives on the boundary will be expressed as a finite difference with values beyond the boundary as 0.}
              \label{fig:5}
            \end{figure}
            \begin{itemize}
                \item On x=$\frac{-1}{2} ( -\uvec{y}) : y \in [-1,1]$
                
                $\nabla \psi \cdot (\Vec{dl}\cross \uvec{z})$ = $-\frac{\partial \psi}{\partial x}|_{(\frac{-1}{2},y)}$
                
                \item On x=$\frac{1}{2}$ ($ +\uvec{y}) : y \in [-1,1]$
                
                $\nabla \psi \cdot (\Vec{dl}\cross \uvec{z})$ = $+\frac{\partial \psi}{\partial x}|_{(\frac{1}{2},y)}$
                
                \item On y=-1 ($+\uvec{x}) : x \in [\frac{-1}{2},\frac{1}{2}]$
                
                $\nabla \psi \cdot (\Vec{dl}\cross \uvec{z})$ = $-\frac{\partial \psi}{\partial y}|_{(x,-1)}$
                
                \item On y=+1 ($-\uvec{x}) : x \in [\frac{-1}{2},\frac{1}{2}]$
                
                $\nabla \psi \cdot (\Vec{dl}\cross \uvec{z})$ = $+\frac{\partial \psi}{\partial y}|_{(x,1)}$
            \end{itemize}
            \vspace{1mm}
            Then :\textemdash
            \vspace{1mm}
            \begin{equation}
                \oint_{L_1} {\grad{\psi}} \cdot \overrightarrow{dl}\times \uvec{z}= -\int^{-1}_{1} \frac{\partial \psi}{\partial x}|_{(\frac{-1}{2},y)} dy-\int^{\frac{1}{2}}_{-\frac{1}{2}}\frac{\partial \psi}{\partial y}|_{(x,-1)} dx+\int^{1}_{-1}\frac{\partial \psi}{\partial x}|_{(\frac{1}{2},y)}+\int^{-\frac{1}{2}}_{\frac{1}{2}}\frac{\partial \psi}{\partial y}|_{(x,1)} \hspace{2 mm}. 
                \label{eqn:21}
            \end{equation}
            \vspace{2mm}
        
            $L_1$ covers the entire region where $B\neq 0$  and thus, the integral will be equal to the total flux.
            \vspace{2mm}
            
                \item Loop 2 : x=$\frac{-1}{2}:y \in [\frac{-7}{8},\frac{7}{8}]             (-\uvec{y}) \cup x= \frac{1}{2}:y \in                                   [\frac{-7}{8},\frac{7}{8}] (+\uvec{y}) \cup y=\frac{-7}{8}:x          \in [\frac{-1}{2},\frac{1}{2}] (+\uvec{x}) \cup y=\frac{7}{8}$           $x\in [\frac{-1}{2},\frac{1}{2}] (-\uvec{x})$
                
                \begin{figure}[h]
                  \centering
                  \includegraphics[width=8 cm]{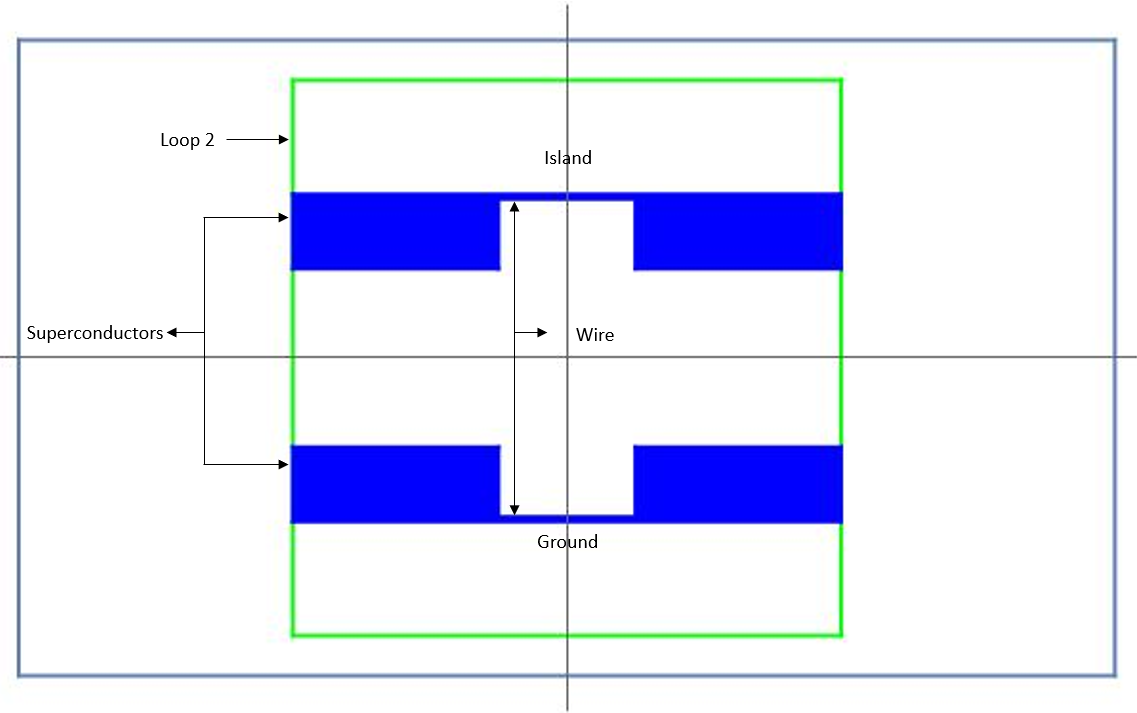}
                  \qquad
                  \caption{\small $L_2$ is a symmetrically defined loop contained entirely within the solution space.}
                  \label{fig:6}
                \end{figure}
                \begin{itemize}
                  \item On x=$\frac{-1}{2} ( -\uvec{y}) : y \in [\frac{-7}{8},\frac{7}{8}]$ 
                  
                  $\nabla \psi \cdot (\Vec{dl}\cross \uvec{z})$ = $-\frac{\partial \psi}{\partial x}|_{(\frac{-1}{2},y)}$
                  
                  \item On x=$\frac{1}{2}$ $( +\uvec{y}) : y \in [\frac{-7}{8},\frac{7}{8}]$
                  
                  $\nabla \psi \cdot (\Vec{dl}\cross \uvec{z})$ = $+\frac{\partial \psi}{\partial x}|_{(\frac{1}{2},y)}$
                  
                  \item On y=$\frac{-7}{8}$ $(+\uvec{x}) : x \in [\frac{-1}{2},\frac{1}{2}]$
                  
                  $\nabla \psi \cdot (\Vec{dl}\cross \uvec{z})$ = $-\frac{\partial \psi}{\partial y}|_{(x,-\frac{7}{8})}$
                  
                  \item On y=$\frac{7}{8} (-\uvec{x}) : x \in [\frac{-1}{2},\frac{1}{2}]$
                  
                  $\nabla \psi \cdot (\Vec{dl}\cross \uvec{z})$ = $+\frac{\partial \psi}{\partial y}|_{(x,\frac{7}{8})}$
                \end{itemize}
                Then : \textemdash
                \vspace{1mm}
                \begin{equation}
                  \oint_{L_2} {\grad{\psi}} \cdot \overrightarrow{dl}\times \uvec{z}= -\int^{-\frac{7}{8}}_{\frac{7}{8}} \frac{\partial \psi}{\partial x}|_{(\frac{-1}{2},y)} dy-\int^{\frac{5}{8}}_{-\frac{1}{2}}\frac{\partial \psi}{\partial y}|_{(x,-\frac{7}{8})} dx+\int^{\frac{7}{8}}_{-\frac{7}{8}}\frac{\partial \psi}{\partial x}|_{(\frac{5}{8},y)} dy+\int^{-\frac{1}{2}}_{\frac{5}{8}}\frac{\partial \psi}{\partial y}|_{(x,\frac{7}{8})} dx \hspace{2 mm}.  
                  \label{eqn:23}
                \end{equation}
                \\
$L_2$ covers the entire region where $B\neq 0$  and thus, the integral will be equal to the total flux.   
\item Loop 3 : x=$\frac{-1}{2}:y \in [\frac{-7}{8},\frac{7}{8}]             (-\uvec{y}) \cup x= \frac{5}{8}:y \in                                   [\frac{-7}{8},\frac{7}{8}] (+\uvec{y}) \cup y=\frac{-7}{8}:x          \in [\frac{-1}{2},\frac{5}{8}] (+\uvec{x}) \cup y=\frac{7}{8}$           $x\in [\frac{-1}{2},\frac{5}{8}] (-\uvec{x})$ 
        
              \begin{figure}[h]
                \centering
                \includegraphics[width=8 cm]{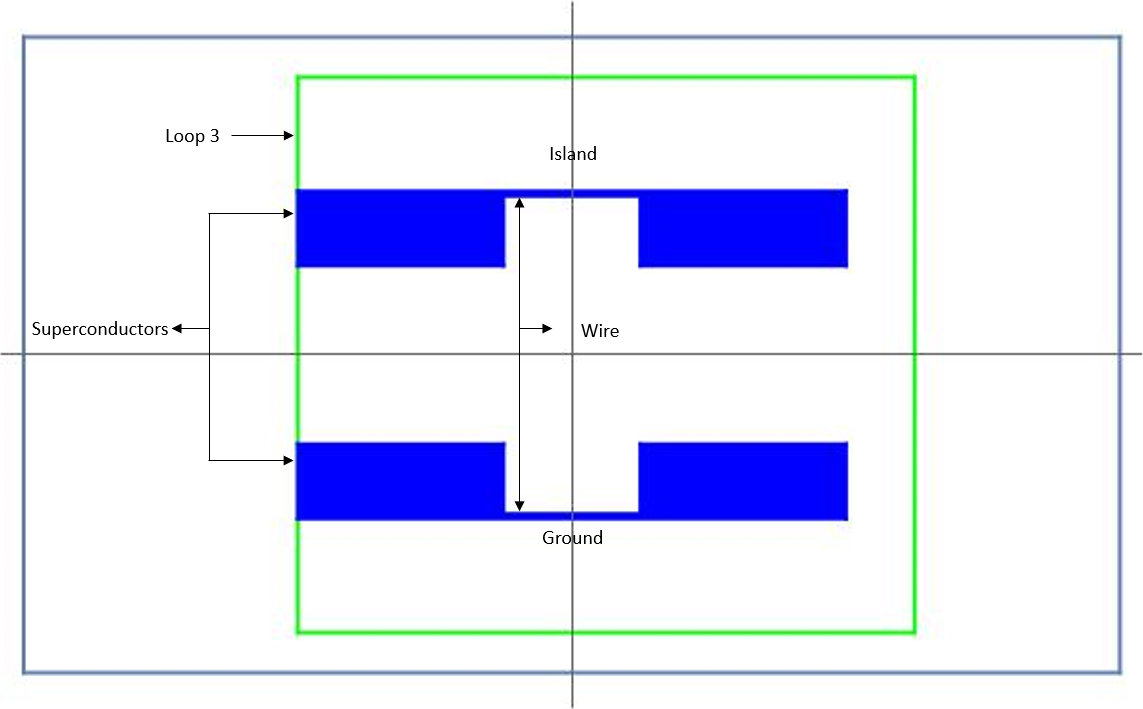}
                \qquad
                \caption{\small $L_3$ is an asymmetrically defined loop. This ensures that the loop integral checks are not satisfied only due to the symmetric nature of the problem in certain instances.}
                \label{fig:7}
               \end{figure}
               \vspace{1mm}
               \begin{itemize}
                  \item On x=$\frac{-1}{2} ( -\uvec{y}) : y \in [\frac{-7}{8},\frac{7}{8}]$ 
                  
                  $\nabla \psi \cdot (\Vec{dl}\cross \uvec{z})$ = $-\frac{\partial \psi}{\partial x}|_{(\frac{-1}{2},y)}$
                  
                  \item On x=$\frac{5}{8}$ $( +\uvec{y}) : y \in [\frac{-7}{8},\frac{7}{8}]$
                  
                  $\nabla \psi \cdot (\Vec{dl}\cross \uvec{z})$ = $+\frac{\partial \psi}{\partial x}|_{(\frac{5}{8},y)}$
                  \item On y=$\frac{-7}{8}$ $(+\uvec{x}) : x \in [\frac{-1}{2},\frac{5}{8}]$
                  $\nabla \psi \cdot (\Vec{dl}\cross \uvec{z})$ = $-\frac{\partial \psi}{\partial y}|_{(x,-\frac{7}{8})}$
                  \item On y=$\frac{7}{8} (-\uvec{x}) : x \in [\frac{-1}{2},\frac{5}{8}]$
                  $\nabla \psi \cdot (\Vec{dl}\cross \uvec{z})$ = $+\frac{\partial \psi}{\partial y}|_{(x,\frac{7}{8})}$
                \end{itemize}
                \vspace{1mm}
                Then :-
                \vspace{1mm}
                \begin{equation}
                  \oint_{L_3} {\grad{\psi}} \cdot \overrightarrow{dl}\times \uvec{z}= -\int^{-\frac{7}{8}}_{\frac{7}{8}} \frac{\partial \psi}{\partial x}|_{(\frac{-1}{2},y)} dy-\int^{\frac{1}{2}}_{-\frac{1}{2}}\frac{\partial \psi}{\partial y}|_{(x,-\frac{7}{8})} dx+\int^{\frac{7}{8}}_{-\frac{7}{8}}\frac{\partial \psi}{\partial x}|_{(\frac{1}{2},y)} dy+\int^{-\frac{1}{2}}_{\frac{1}{2}}\frac{\partial \psi}{\partial y}|_{(x,\frac{7}{8})} dx \hspace{2 mm}.
                  \label{eqn:22}
               \end{equation}
               \vspace{1mm}
               $L_3$ covers the entire region where $B\neq 0$  and thus, the integral will be equal to the total flux.
        \end{enumerate}

\vspace{2mm}
\flushleft For each of the simulations, these loop integrals are evaluated and their equality to one-another is checked. These loop integrals are found to all be equal for all the simulations performed - even if they deviate from the theoretical ideal. This ensures that the numerical analysis is self-consistent.
\vspace{2mm}
\\
The deviation of the values of these integrals from the theoretical ideal case is due to the effect of the discreteness of the grid on the magnetic field definition. This deviation can be minimized, without using finer grids and more computation, by defining the field on the `plaquettes' via line integrals over each square in the grid - rather than associating the field with the lattice points. However, since $\alpha_1,\alpha_2$ are calculated as line integrals themselves, normalized by the total flux, this error should not greatly alter the values we are interested in. This assumption is validated by the second set of checks performed.
\vspace{1mm}
\subsection{Finer grid tests}
To ensure that errors due to the discreteness of the grid \textbf{do not} affect the ratio of the $\alpha$ values beyond an accepted tolerance limit, a select set of simulations were run with on a finer 256 $\cross$ 256 grid. The change in the $\alpha$ values, and their ratio, on using a finer grid is used to obtain an estimate for an `acceptable' tolerance limit that characterizes the accuracy of the simulations.
\\
The acceptable tolerance limit in the ratio of the two $\alpha$ values, and thus also the ratio of the two capacitances, is found to be of the order of $10^{-2}$ - capped at $\approx 4*10^{-2}$, which is also of the order of the standard truncation error estimate of \textit{O(\textbf{h})} (`h' being the grid spacing) for a finite difference system whose solution is only $C^2$ continuous and given that $\alpha$ is calculated as an integral of directional derivatives, their error will be of the same order.
\vspace{2mm}
\subsection{Tests on the finite boundary approximation}
To determine to what extent the `finitude' of $\infty$ affects the solution, we can increase the size of the solution space $\{x_\text{min},x_\text{max}\} \& \{y_\text{min},y_\text{max}\}$ and examine how the altering of the zero rest condition affects the solution.
\vspace{2mm}
\\
As the size of the solution space scales, so should the grid sizing, so as to maintain the same number of grid points for the problem structure to obtain a good enough comparison of the manner in which the finitude of the space affects the solution.
\vspace{2mm}
\\
It is found that the solution does not change by much on increasing the solution space and that the variation is well within the `acceptable' tolerance limit as obtained through the finer grid tests.
\vspace{2mm}
\\
The amount by which most of the values change as we alter the space from $\left[-1,1\right] \otimes \left[-1,1\right]$ to $\left[-2,2\right] \otimes \left[-2,2\right]$ and from $\left[-2,2\right] \otimes \left[-2,2\right]$ to $\left[-4,4\right] \otimes \left[-4,4\right]$ is $\approx 2*10^{-3}$, which is equal to the difference between the values of the loop integrals we have obtained and the limit as the number of divisions $\to \infty$ in both directions.
\vspace{2mm}
\\
Thus, for the kind of accuracy we require, the finitude approximation is `good enough' and we can perform all the required simulations on a 128 x 128 grid on the solution space - $\left[-1,1\right] \otimes \left[-1,1\right]$
\maketitle
\section{Results in accordance with the predictions of You et al.}
\vspace{2 mm}
\label{sec:12}
The prediction of Ref.~\hyperref[sec:Ref]{[2]} that $\alpha={C_2}/{C_\text{tot}}$ is \textbf{not} violated for a large set of spatial positions of the magnetic field. 
\\
In settings in which the flux is concentrated in a region at the centre of the two superconductors and distributed such that reflection symmetry about the two axes holds, it is found that the predictions of Ref.~\hyperref[sec:Ref]{[2]} hold precisely. Results that do \textbf{not} signal a departure from Ref.~\hyperref[sec:Ref]{[2]} are stated in this section.
\vspace{2 mm}
\subsection{Structural specification}
\vspace{1mm}
\begin{enumerate}
                    \item $\{ \text{B}_\text{xextentn},\text{B}_\text{xextentp} \}$ $\rightarrow$ $\{ \frac{-1}{16},\frac{1}{16}\}$
                    \item $\{ \text{B}_\text{yextentn},\text{B}_\text{yextentp} \}$ $\rightarrow$ $\{ \frac{-1}{16},\frac{1}{16}\}$
                    \item lsc $\rightarrow \frac{1}{8}$
                    \item d $\rightarrow \frac{7}{16}$
                    \item pos $\rightarrow \frac{1}{2}$ 
                    \item t $\rightarrow \frac{1}{64}$
\end{enumerate}
\vspace{2 mm}
\subsection{Results}
Since the flux in the gap between the two plates is zero, $\alpha_1$ and $\alpha_2$ are independent of the exact paths between the two plates we choose for their evaluation.
\\
We compute a line integral of $A_\text{irr}$ over a loop $\left( L_4 \right)$ that contains within it the paths chosen for the computation of $\alpha_1$ and $\alpha_2$, completed within the bulk of the superconductor. Given that the stream function is constant within the bulk, the only contribution to this loop integral will be via the paths used to calculate the $\alpha$ values. Thus, by definition, $\alpha_1$ and $\alpha_2$ should sum to give the integral over $L_4$. For all these computations, the loop integrals in Eq.~(\hyperref[eqn:21]{21}),Eq.~(\hyperref[eqn:22]{22}) and Eq.~(\hyperref[eqn:23]{23}) all evaluate to $\approx 0.197$ - within the acceptable error limit.
\newpage
\begin{enumerate}
            \item $w_\text{r}=\frac{1}{16}$
                 
                 \vspace{2 mm}
                 
                 \begin{tabular}{||c c c c c c c c c c c c||}
                 \hline
                  $w_\text{l}$ & $L_4$ & $\alpha_1$ SM & $\alpha_1$ IM & $\alpha_2$ SM & $\alpha_2$ IM & $\frac{\alpha_1}{\alpha_2}$ SM & $\frac{\alpha_1}{\alpha_2}$ IM & $\frac{C_2}{C_1}$ SM$_0$ & $\frac{C_2}{C_1}$ IM$_0$ & $\frac{C_2}{C_1}$ SM$_\text{lsc}$ & $\frac{C_2}{C_1}$ IM$_\text{lsc}$ \\ [0.5ex]         
                  \hline \hline
                  $\frac{1}{16}$ & 0.0218 & 0.0109 & 0.0126 & 0.0109 & 0.0126 & 1 & 1 & 1 & 0.977 & 1 & 1\\ 
                   \hline
                  $\frac{2}{16}$ & 0.0218 & 0.00970 & 0.0111 & 0.0121 & 0.0140 & 0.799 & 0.792 & 0.819 & 0.790 & 0.810 & 0.788 \\
                   \hline
                  $\frac{3}{16}$ & 0.0217 & 0.00873 & 0.00999 & 0.0131 & 0.0152 & 0.667 & 0.659 & 0.694 & 0.664 & 0.681 & 0.652 \\
                   \hline
                  $\frac{4}{16}$ & 0.0216 & 0.00795 & 0.00909 & 0.0139 & 0.0161 & 0.551 & 0.572 & 0.601 & 0.572 & 0.588 & 0.556 \\
                   \hline
                  $\frac{5}{16}$ & 0.0215 & 0.00730 & 0.00834 & 0.0145 & 0.0168 & 0.502 & 0.496 & 0.530 & 0.502 & 0.517 & 0.485 \\
                  \hline
                  \hline
                  \end{tabular}
                  \vspace{4 mm}
                  \item $w_\text{r}=\frac{2}{16}$
                 
                 \begin{tabular}{||c c c c c c c c c c c c||}
                 \hline
                  $w_\text{l}$ & $L_4$ & $\alpha_1$ SM & $\alpha_1$ IM & $\alpha_2$ SM & $\alpha_2$ IM & $\frac{\alpha_1}{\alpha_2}$ SM & $\frac{\alpha_1}{\alpha_2}$ IM & $\frac{C_2}{C_1}$ SM$_0$ & $\frac{C_2}{C_1}$ IM$_0$ & $\frac{C_2}{C_1}$ SM$_\text{lsc}$ & $\frac{C_2}{C_1}$ IM$_\text{lsc}$ \\ [0.5ex]          
                  \hline \hline
                  $\frac{1}{16}$ & 0.0218 & 0.0121 & 0.0140 & 0.0097 & 0.0111 & 1.251 & 1.265 & 1.221 & 1.214 & 1.235 & 1.268 \\ 
                   \hline
                  $\frac{2}{16}$ & 0.0218 & 0.0109 & 0.0125 & 0.0109 & 0.0125 & 1 & 1 & 1 & 1 & 0.982 & 1 \\
                   \hline
                  $\frac{3}{16}$ & 0.0218 & 0.0099 & 0.0113 & 0.0119 & 0.0136 & 0.834 & 0.833 & 0.874 & 0.824 & 0.842 & 0.827 \\
                   \hline
                  $\frac{4}{16}$ & 0.0217 & 0.0091 & 0.0104 & 0.0127 & 0.0154 & 0.716 & 0.715 & 0.734 & 0.710 & 0.727 & 0.706 \\
                   \hline
                  $\frac{5}{16}$ & 0.0217 & 0.0084 & 0.0096 & 0.0134 & 0.0149 & 0.628 & 0.627 & 0.648 & 0.623 & 0.639 & 0.615 \\
                  \hline
                  \hline
                  \end{tabular}
                  
                 \vspace{2 mm}
                 \item $w_\text{r}=\frac{3}{16}$
                 
                 \vspace{2 mm}
                
                 \begin{tabular}{||c c c c c c c c c c c c||}
                 \hline
                  $w_\text{l}$ & $L_4$ & $\alpha_1$ SM & $\alpha_1$ IM & $\alpha_2$ SM & $\alpha_2$ IM & $\frac{\alpha_1}{\alpha_2}$ SM & $\frac{\alpha_1}{\alpha_2}$ IM & $\frac{C_2}{C_1}$ SM$_0$ & $\frac{C_2}{C_1}$ IM$_0$ & $\frac{C_2}{C_1}$ SM$_\text{lsc}$ & $\frac{C_2}{C_1}$ IM$_\text{lsc}$ \\ [0.5ex]         
                  \hline \hline
                  $\frac{1}{16}$ & 0.0217 & 0.0131 & 0.0152 & 0.00874 & 0.0099 & 1.519 & 1.500 & 1.441 & 1.450 & 1.467 & 1.533 \\ 
                   \hline
                  $\frac{2}{16}$ & 0.0218 & 0.0119 & 0.0136 & 0.0099 & 0.0113 & 1.199 & 1.201 & 1.181 & 1.172 & 1.188 & 1.209 \\
                   \hline
                  $\frac{3}{16}$ & 0.0218 & 0.0109 & 0.0125 & 0.0109 & 0.0125 & 1 & 1 & 1 & 0.984 & 1 & 1 \\
                   \hline
                  $\frac{4}{16}$ & 0.0218 & 0.0101 & 0.0115 & 0.0117 & 0.0134 & 0.859 & 0.858 & 0.867 & 0.848 & 0.863 & 0.853 \\
                   \hline
                  $\frac{5}{16}$ & 0.0218 & 0.00937 & 0.0107 & 0.0124 & 0.0142 &  0.753 & 0.753 & 0.765 & 0.744 & 0.759 & 0.743 \\
                  \hline
                  \hline
                  \end{tabular}

                  \vspace{2 mm}
                  \item $w_\text{r}=\frac{4}{16}$

                 \begin{tabular}{||c c c c c c c c c c c c||}
                 \hline
                  $w_\text{l}$ & $L_4$ & $\alpha_1$ SM & $\alpha_1$ IM & $\alpha_2$ SM & $\alpha_2$ IM & $\frac{\alpha_1}{\alpha_2}$ SM & $\frac{\alpha_1}{\alpha_2}$ IM & $\frac{C_2}{C_1}$ SM$_0$ & $\frac{C_2}{C_1}$ IM$_0$ & $\frac{C_2}{C_1}$ SM$_\text{lsc}$ & $\frac{C_2}{C_1}$ IM$_\text{lsc}$ \\ [0.5ex]       
                  \hline \hline
                  $\frac{1}{16}$ & 0.0216 & 0.0139 & 0.0161 & 0.0091 & 0.0085 & 1.747 & 1.770 & 1.663 & 1.686 & 1.700 & 1.797 \\ 
                   \hline
                  $\frac{2}{16}$ & 0.0217 & 0.0127 & 0.0146 & 0.0091 & 0.0104 & 1.399 & 1.396 & 1.362 & 1.364 & 1.376 & 1.417\\
                   \hline
                  $\frac{3}{16}$ & 0.0218 & 0.0117 & 0.0134 & 0.0101 & 0.0115 & 1.165 & 1.165 & 1.153 & 1.145 & 1.158 & 1.172 \\
                   \hline
                  $\frac{4}{16}$ & 0.0218 & 0.0109 & 0.0125 & 0.0109 & 0.0125  & 1 & 1 & 1 & 0.987 & 1 & 1 \\
                   \hline
                  $\frac{5}{16}$ & 0.0218 & 0.0102 & 0.0116 & 0.0116 & 0.0133 & 0.877 & 0.877 & 0.882 & 0.866 & 0.879 & 0.871 \\
                  \hline
                  \hline
                  \end{tabular}

                  \vspace{2 mm}
                  
                  \item $w_\text{r}=\frac{5}{16}$
                 
                 \vspace{2 mm}
                 
                 \begin{tabular}{||c c c c c c c c c c c c||}
                 \hline
                  $w_\text{l}$ & $L_4$ & $\alpha_1$ SM & $\alpha_1$ IM & $\alpha_2$ SM & $\alpha_2$ IM & $\frac{\alpha_1}{\alpha_2}$ SM & $\frac{\alpha_1}{\alpha_2}$ IM & $\frac{C_2}{C_1}$ SM$_0$ & $\frac{C_2}{C_1}$ IM$_0$ & $\frac{C_2}{C_1}$ SM$_\text{lsc}$ & $\frac{C_2}{C_1}$ IM$_\text{lsc}$ \\ [0.5ex]         
                  \hline \hline
                  $\frac{1}{16}$ & 0.0215 & 0.0145 & 0.0168 & 0.00730 & 0.00834 & 1.992 & 2.018 & 1.885 & 1.924 & 1.933 & 2.062 \\ 
                   \hline
                  $\frac{2}{16}$ & 0.0217 & 0.0134 & 0.0154 & 0.00834 & 0.00962 & 1.592 & 1.596 & 1.544 & 1.556 & 1.565 & 1.626 \\
                   \hline
                  $\frac{3}{16}$ & 0.0218 & 0.0124 & 0.0142 & 0.00937 & 0.0107 & 1.328 & 1.328 & 1.308 & 1.307 & 1.317 & 1.345 \\
                   \hline
                  $\frac{4}{16}$ & 0.0218 & 0.0116 & 0.0133 & 0.0102 & 0.0116  & 1.140 & 1.140 & 1.134 & 1.126 & 1.137 &  1.148 \\
                   \hline
                  $\frac{5}{16}$ & 0.0218 & 0.0109 & 0.0125 & 0.0109 & 0.0125 & 1 & 1 & 1 & 0.988 & 1 & 1 \\
                  \hline
                  \hline
                  \end{tabular}
        \end{enumerate}

In all of these results, the flux is \textbf{spatially uniform}, concentrated at the centre of the structure and is symmetric on reflection in the two axes.
\vspace{2mm}
\\
Scaling the magnetic field without altering its spatial distribution only leads to a scaling of the solution, due to the nature of the Poisson equation. Thus, if the flux varies continuously in time as a scaling alone under the constraints above, without spatial variation, the predictions of Ref.~\hyperref[sec:Ref]{2} are \textbf{guaranteed} to hold and the capacitive elements introduced in moving from the continuous geometry in Fig.~\hyperref[fig:1]{1} to the lumped element equivalent in Fig.~\hyperref[fig:2]{2} will be the \textbf{electrostatic junction capacitances}. 
\vspace{2 mm}
\\
\textbf{Note:\textemdash} $C_2/C_1$ IM$_0$ $\approx C_2/C_1$ IM$_\text{lsc}$, thus supporting the fact that there is negligible charge on the thin wire and ensuring that the two delimitations of capacitances are almost equivalent. Also, the decreasing ${C_2}/{C_1}$ values on increasing $w_\text{l}$ is suggestive of the fact that the capacitances increase with an increase in area - similar to the behaviour seen in isolated parallel-plate capacitances.
\vspace{2mm}
\\
Since we are \textbf{not} concerned with setting out a set of necessary or sufficient conditions for the predictions of Ref.~\hyperref[sec:Ref]{2} to hold - and doing so via numerical methods will in fact be farcical, this set of results suffices to show that there are settings in which they do hold. The following section indicates that there are settings even in a geometry as straightforward as that in Fig.~\hyperref[fig:1]{1}, capacitances that depend on the spatial distribution and nature of the flux and which are negative and even singular, may appear during discretization. 
\maketitle
\section{Results contrary to the predictions of You et al.}
Consider situations in which the magnetic flux is no longer distributed in a symmetric manner or concentrated at the centre of the structure. It is found in the following setups that the ratio \textemdash $\alpha_1/\alpha_2$, is no longer equal to the ratio of the electrostatic junction capacitances \textemdash $C_2/C_1$. We only tabulate the ratios \textemdash $\alpha_1/\alpha_2$ and $C_2/C_1$ as we only need to compare those.
\subsection{Structural specification}
\begin{enumerate}
                    \item lsc $\rightarrow \frac{1}{8}$
                    \item pos $\rightarrow \frac{1}{4}$ - The y-coordinate that specifies the position locale of the superconductor
                    \item t $\rightarrow \frac{1}{64}$
                    \item d $\rightarrow \frac{14}{64}$ - so that the gap between the upper and the lower superconductors will be $\frac{1}{32}$
                    \item $w_\text{l}$ = $\frac{3}{8}$
                    \item $w_\text{r}$ = $\frac{3}{8}$
\end{enumerate}
As expected, $\frac{C_2}{C_1} = 1$.
\subsection{Results}
\vspace{2mm}
\subsubsection{Varying the spatial extents of the magnetic field}
\vspace{2 mm}
\label{sec:13.2.1}
We vary the spatial extents of the magnetic field, while keeping the total flux in Eq.~(\hyperref[eqn:1]{1}) constant by constraining the entire field to lie in the region enclosed by the two Josephson junctions, within the `U-shaped' region in Fig.~\hyperref[fig:1]{1}.

           \begin{enumerate}
                \item $\{B_\text{yextentn},B_\text{yextentp}\} \rightarrow \{\frac{-1}{16},\frac{1}{16}\}$
                \vspace{2 mm}
               
               \begin{tabular}{||c c c c c c c c||}
                  \hline
                  $B_\text{xextentn}$ & $B_\text{xextentp}$ & $\frac{\alpha_1}{\alpha_2}$ SM & $\frac{\alpha_1}{\alpha_2}$ IM & $\frac{C_2}{C_1}$ SM$_0$ & $\frac{C_2}{C_1}$ IM$_0$ & $\frac{C_2}{C_1}$ SM$_{lsc}$ & $\frac{C_2}{C_1}$ IM$_{lsc}$ \\ [1.0  ex]     
                  
                  \hline \hline
                  0 & $\frac{1}{64}$ & 0.995 & 0.995 & 1 & 0.994 & 1 & 1 \\ [1.0 ex]
                   \hline
                  $\frac{1}{64}$ & $\frac{2}{64}$ & 0.985 & 0.985 & 1 & 0.994 & 1 & 1 \\ [1.0 ex]
                   \hline
                   $\frac{2}{64}$ & $\frac{3}{64}$ & 0.975 & 0.975 & 1 & 0.994 & 1 & 1 \\ [1.0 ex]
                   \hline
                  $\frac{3}{64}$ & $\frac{4}{64}$ & 0.964 & 0.964 & 1 & 0.994 & 1 & 1 \\ [1.0 ex]
                  \hline
                  $\frac{4}{64}$ & $\frac{5}{64}$ & 0.953 & 0.953 & 1 & 0.994 & 1 & 1 \\ [1.0 ex]
                  \hline
                  $\frac{5}{64}$ & $\frac{6}{64}$ & 0.941 & 0.941 & 1 & 0.994 & 1 & 1 \\ [1.0 ex]
                  \hline
                  $\frac{6}{64}$ & $\frac{7}{64}$ & 0.927 & 0.927 & 1 & 0.994 & 1 & 1 \\ [1.0 ex]
                  \hline
                  \hline
                \end{tabular}
            \item $\{B_\text{yextentn},B_\text{yextentp}\} \rightarrow \{0,\frac{1}{16}\}$
            
            \vspace{2 mm}
            
               \begin{tabular}{||c c c c c c c c||}
                  \hline
                  $B_\text{xextentn}$ & $B_\text{xextentp}$ & $\frac{\alpha_1}{\alpha_2}$ SM & $\frac{\alpha_1}{\alpha_2}$ IM & $\frac{C_2}{C_1}$ SM$_0$ & $\frac{C_2}{C_1}$ IM$_0$ & $\frac{C_2}{C_1}$ SM$_{lsc}$ & $\frac{C_2}{C_1}$ IM$_{lsc}$ \\ [1.0  ex]     
                  
                  \hline \hline
                  0 & $\frac{1}{64}$ & 0.995 & 0.995 & 1 & 0.994 & 1 & 1 \\ [1.0 ex]
                   \hline
                  $\frac{1}{64}$ & $\frac{2}{64}$ & 0.985 & 0.985 & 1 & 0.994 & 1 & 1 \\ [1.0 ex]
                   \hline
                   $\frac{2}{64}$ & $\frac{3}{64}$ & 0.975 & 0.975 & 1 & 0.994 & 1 & 1 \\ [1.0 ex]
                   \hline
                  $\frac{3}{64}$ & $\frac{4}{64}$ & 0.963 & 0.963 & 1 & 0.994 & 1 & 1 \\ [1.0 ex]
                  \hline
                  $\frac{4}{64}$ & $\frac{5}{64}$ & 0.951 & 0.951 & 1 & 0.994 & 1 & 1 \\ [1.0 ex]
                  \hline
                  $\frac{5}{64}$ & $\frac{6}{64}$ & 0.939 & 0.939 & 1 & 0.994 & 1 & 1 \\ [1.0 ex]
                  \hline
                  $\frac{6}{64}$ & $\frac{7}{64}$ & 0.924 & 0.924 & 1 & 0.994 & 1 & 1 \\ [1.0 ex]
                  \hline
                  \hline
                \end{tabular}
                
                \vspace{2 mm}
            \newpage    
            \item $\{B_\text{yextentn},B_\text{yextentp}\} \rightarrow \{0,0\}$
            
            \vspace{2 mm}
            
               \begin{tabular}{||c c c c c c c c||}
                  \hline
                  $B_{xextentn}$ & $B_{xextentp}$ & $\frac{\alpha_1}{\alpha_2}$ SM & $\frac{\alpha_1}{\alpha_2}$ IM & $\frac{C_2}{C_1}$ SM$_0$ & $\frac{C_2}{C_1}$ IM$_0$ & $\frac{C_2}{C_1}$ SM$_{lsc}$ & $\frac{C_2}{C_1}$ IM$_{lsc}$ \\ [1.0  ex]     
                  
                  \hline \hline
                  0 & $\frac{1}{64}$ & 0.994 & 0.994 & 1 & 0.994 & 1 & 1 \\ [1.0 ex]
                   \hline
                  $\frac{1}{64}$ & $\frac{2}{64}$ & 0.983 & 0.983 & 1 & 0.994 & 1 & 1 \\ [1.0 ex]
                   \hline
                   $\frac{2}{64}$ & $\frac{3}{64}$ & 0.971 & 0.971 & 1 & 0.994 & 1 & 1 \\ [1.0 ex]
                   \hline
                  $\frac{3}{64}$ & $\frac{4}{64}$ & 0.958 & 0.958 & 1 & 0.994 & 1 & 1 \\ [1.0 ex]
                  \hline
                  $\frac{4}{64}$ & $\frac{5}{64}$ & 0.942 & 0.942 & 1 & 0.994 & 1 & 1 \\ [1.0 ex]
                  \hline
                  $\frac{5}{64}$ & $\frac{6}{64}$ & 0.924 & 0.924 & 1 & 0.994 & 1 & 1 \\ [1.0 ex]
                  \hline
                  $\frac{6}{64}$ & $\frac{7}{64}$ & 0.904 & 0.904 & 1 & 0.994 & 1 & 1 \\ [1.0 ex]
                  \hline
                  \hline
                \end{tabular}
            \end{enumerate}
Thus, the ratio - $\alpha_1/\alpha_2$ deviates from the values predicted by Ref.~\hyperref[sec:Ref]{[2]} in that it is \textbf{not} equal to the ratio of the \textbf{electrostatic junction capacitances}. Moreover, if the magnetic field varies temporally such that its spatial extents change over time, the capacitances which appear in the lumped element must thus be \textbf{time-dependent}. This proves that the capacitive values are not \textbf{only} dependent on the geometry of the continuous structure but also the spatial distribution and the nature of the field - in accordance with the predictions of Ref.~\hyperref[sec:Ref]{[1]}.
\vspace{2 mm}

\textbf{Note :\textemdash}The deviation from the predictions of Ref.~\hyperref[sec:Ref]{[2]} is not large in that the maximum deviation we have obtained is $\approx 0.1$ is due to the dimensions of the continuous geometry we have considered and larger deviations can be obtained on increasing the size of the `U-shaped' region. A few tabulations of the deviations from the predictions of You et al. for larger dimensions of the `U-shaped' region have been provided in the appendix. At the moment, it suffices to show that deviations from the predictions of You et al. are evident even when the field is confined to the `U-shaped' region. The deviation seen in the tabulated values is suggestive of the fact that moving the field extents closer to either of the `capacitances', while ensuring that the total flux remains constant, increases the deviation from the predictions of You et al.
\vspace{2mm}
\\
We have already shown that the predictions closely mirror the resulting output as long as the flux is confined to the centre of the `U-shaped' region and symmetry on reflection in the line x=0 is maintained. Thus, it is to be expected that breaking this symmetry and increasing the `extent' of this asymmetry by altering the extents of the magnetic flux, results in increasing deviations from the predictions made by You et al.
\vspace{1mm}
\subsubsection{Placing the magnetic field between the capacitive plates}
\begin{figure}[h]
    \centering
    \includegraphics[width=8 cm]{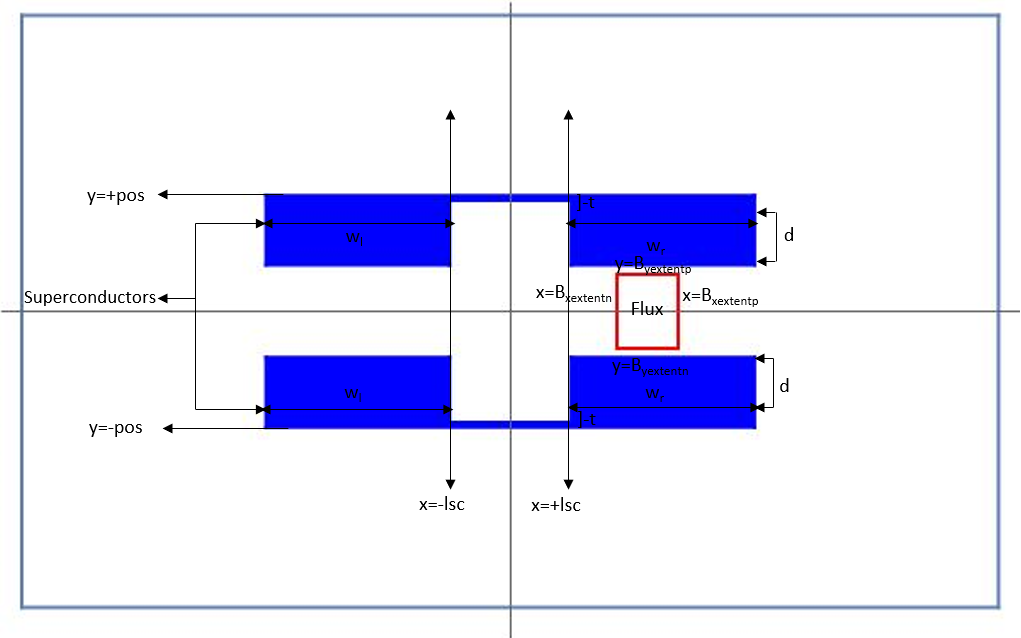}
    
    \qquad
    \caption{\small The magnetic field is placed in the gap between two plates of the capacitor on the right \textemdash $C_2$. In these cases, we are required to have negative, and potentially singular capacitances. While tabulating the $\alpha$ values, we consider the maximum and minimum possible values of $\alpha_2$, which will now depend on the position of the second Josephson junction. Since $\alpha_1$ is fixed for a given driving field, and $\alpha_1 + \alpha_2 = \Phi$, the flux enclosed by the two Josephson junctions, as the position of the second Josephson junction passes through the region where the flux is concentrated, left to right, the value of $\alpha_2$ will increase and thus, the minimum value of $\alpha_2$ will be when the second Josephson junction is at the left extreme, between the gaps of the two plates, and the maximum value of $\alpha_2$ will be when it is beyond the region where the flux is concentrated \textemdash on the right.}
    \label{fig:8}
\end{figure}

\vspace{2 mm}
\label{sec:13.2.2}
We now place the entire magnetic field in the gap between two plates of one of the capacitors in the structure. In such situations, the position where the Josephson junction is placed affects the phase that enters Eq.~(\hyperref[eqn:2]{2}) due to the flux enclosed in the area by the Josephson junctions as Eq.~(\hyperref[eqn:1]{1}), which can be seen as the integrals - Eq.~(\hyperref[eqn:13]{13}),Eq.~(\hyperref[eqn:14]{14}) - now depending on the paths $J_k$. Since the phases that now appear within the Hamiltonian for the system now depend on the position of the Josephson junction that is connected across the `plates' between which the magnetic field is now positioned, so will the `$\alpha$' values - which essentially determine the weight of the total enclosed flux $\Phi$(t) that is `associated' with each of the Josephson junctions.
\vspace{2mm}
\\
Without any loss of generality, we assume that the field is placed between the second set of `plates' as in Fig.~\hyperref[fig:8]{8}. Thus, the position of the Josephson junction between the plates of `$C_1$' and thus, the location of $J_1$ does not affect the value of $\alpha_1$.nHowever, the choice of position of the Josephson junction across `$C_2$' is important and affects $\alpha_2$. All the positions are valid since in a physical structure, one could place the Josephson junction anywhere across the two plates.
\vspace{2mm}
\\
Moreover, since $\alpha_1 + \alpha_2$ must sum to $\Phi$, the total flux enclosed by the two Josephson junctions, on moving the second Josephson junction through the field between the plates of `$C_2$', $\alpha_2$ will increase linearly. Thus, it suffices to calculate $\alpha_2$ for the two extreme cases - viz. when the total flux enclosed by the two junctions is 0 and when the entire region where the magnetic field is non-zero is enclosed by the two junctions.
\vspace{2mm}
\\
In the case where the second Josephson junction is considered to be placed such that the flux enclosed as $\Phi =0$, note that the RHS of Eq.~(\hyperref[eqn:16]{16}) and Eq.~(\hyperref[eqn:17]{17}) go to zero as $\Phi=0$. However, $\alpha_1$ and $\alpha_2$ are \textbf{non-zero}. This signals the need for singular capacitances in the lumped element circuit drawn up in Fig.~\hyperref[fig:2]{2} \textemdash precisely as stated in Ref.~\hyperref[sec:Ref]{[1]}. However, the total capacitance can still be positive and finite as we are only required to have $C_2/C_\text{tot}$ and $C_1/C_\text{tot}$ to be singular values. Cases where $\alpha_1$ \textbf{or} $\alpha_2$ is negative will thus signal the requirement for negative capacitances as well \textemdash as predicted by Ref.~\hyperref[sec:Ref]{1}. Note that both $\alpha_1$ and $\alpha_2$ being negative, on account of the total flux being negative, can be dealt with using positive capacitances alone.
\vspace{2mm}
\\
\textbf{Note :\textemdash} The flux passing through each Josephson junction must be less than the flux quantum. Since we have not dealt with any units in the analysis performed, we can assume this is true without any special considerations having to be made.
\vspace{2 mm}
\newpage
\begin{enumerate}
                \item $\{B_\text{yextentn},B_\text{yextentp}\}$ $\rightarrow \{\frac{-1}{64},\frac{1}{64}\}$ : Total Flux = $3.92 * 10^{-3}$; $\frac{C_2}{C_1}=1$
                \vspace{2 mm}
                \\
                \begin{tabular}{||c c c c c c c||}
                  \hline
                  $B_\text{xextentn}$ & $B_\text{xextentp}$ & $\alpha_1$ & $\alpha_2$ (min) & $\alpha_2$ (max) & $\frac{\alpha_1}{\alpha_2} \left(min\right)$ IM & $\frac{\alpha_1}{\alpha_2} \left(max\right)$ IM \\ [1.0  ex]     
                  
                  \hline \hline
                  $\frac{2}{8}$ & $\frac{3}{8}$ & -0.00263 & 0.00263 & 0.00754 & -1 & -0.348\\ [1 ex]
                   \hline
                  $\frac{9}{64}$ & $\frac{17}{64}$ & -0.00529 & 0.00529 & 0.0105 & -1 & -0.502\\ [1 ex]
                   \hline
                   $\frac{10}{64}$ & $\frac{18}{64}$ & -0.00333 & 0.00333 & 0.00571 & -1 & -0.583 \\[1 ex]
                   \hline
                   $\frac{11}{64}$ & $\frac{19}{64}$ & -0.00318 & 0.00318 & 0.00586 & -1 & -0.542 \\[1 ex]
                   \hline
                   $\frac{12}{64}$ & $\frac{20}{64}$ & -0.00303 & 0.00303 & 0.00601 & -1 & -0.502 \\[1 ex]
                   \hline
                   $\frac{13}{64}$ & $\frac{21}{64}$ & -0.00287 & 0.00287 & 0.00617 & -1 & -0.465 \\[1 ex]
                   \hline
                   $\frac{14}{64}$ & $\frac{22}{64}$ & -0.00272 & 0.00272 & 0.00632 & -1 & -0.430 \\[1 ex]

                  \hline
                  \hline
                \end{tabular}

            \end{enumerate}
\vspace{2 mm}
Thus the $\alpha$ values and their ratios greatly deviate from that predicted by Ref.~\hyperref[sec:Ref]{[2]}. $\alpha_1$ and $\alpha_2$ being non-zero when $\Phi=0$ implies the need for singular effective capacitances - \textbf{which are not equal} to the junction capacitances, the ratio of which is given by the ratio of the corresponding $\alpha$ values for those positions. The ratio being negative implies that one of the capacitances will have to be assigned a negative value. We choose this to be $C_\text{eff,2}$ so that $C_\text{tot}$ is positive and definite. 
\vspace{4mm}
\hrule
\newpage
\maketitle
\section{Conclusion}
Thus, it is clear to see that $\alpha_1/\alpha_2$ deviates significantly from $C_2/C_1$. This implies that a transition to the discrete setting via lumped element methods requires that we allow the effective capacitances in Fig.~\hyperref[fig:2]{2} to be time-dependent,negative and even potentially, singular \textemdash congruous to the predictions made by Ref.~\hyperref[sec:Ref]{1}.
\vspace{2mm}
\\
A method to avoid this unnatural assignment of values to the effective capacitances is proposed in Ref.~\hyperref[sec:Ref]{1} as a `refined lumped element approach' in which a countable infinity of degrees of freedom are introduced by the introduction of capacitances at \textit{all} nodes. In such an approach, the irrotational gauge procedure is equivalent to that used in Ref.~\hyperref[sec:Ref]{[2]}. The internal workings of the \textit{island} is also captured in this method, in which the island is treated as an infinite transmission line. 
\vspace{2mm}
\\
However, as stated in Ref.~\hyperref[sec:Ref]{1}, the finite degree of freedom approach with two capacitances is still valid as long as we allow for time-dependent, negative and potentially singular capacitance values. This implies that at any given `snapshot' of time, there exists a non-unitary map - non-unitary, owing to the differences in the degrees of freedom - that transitions from the `refined lumped element approach' to the coarse method in Ref.~\hyperref[sec:Ref]{2}, which is characteristically dependent on both the field distribution as well as the structural configuration. Characterizing this map would allow us to analytically determine the capacitance values that must be assigned to the capacitances in Fig.~\hyperref[fig:2]{2} and their variation with time. It remains to be seen if knowledge of this map for a given evolution will allow us to continue with a finite degrees of freedom.
\vspace{10 mm}
\hrule
\maketitle
\vspace{10 mm}
\label{sec:Ref}
\begin{align*}
    [1]&- \text{R.P Riwar and D.P DiVincenzo, Circuit quantization with time-dependent magnetic fields for realistic geometries}
    \\
     &~~\text{ \href{https://arxiv.org/abs/2103.03577}{arXiv pre-print (2021)}}
    \\
    [2] &-\text{X. You, J. A. Sauls, and J. Koch, Circuit quantization in
the presence of time-dependent external flux }
     \\
     &~~\text{ \href{https://journals.aps.org/prb/abstract/10.1103/PhysRevB.99.174512}{Phys. Rev.B 99, 174512 (2019)}}\label{ref:2}
\end{align*}
\newpage
\maketitle
\section{Appendix}
\subsection{Contour plots of the stream function in accordance with predictions in You et al.}
\label{sec:15.2}
\begin{enumerate}
                    \item $\{ \text{B}_\text{xextentn},\text{B}_\text{xextentp} \}$ $\rightarrow$ $\{ \frac{-1}{16},\frac{1}{16}\}$
                    \item $\{ \text{B}_\text{yextentn},\text{B}_\text{yextentp} \}$ $\rightarrow$ $\{ \frac{-1}{16},\frac{1}{16}\}$
                    \item lsc $\rightarrow \frac{1}{8}$
                    \item d $\rightarrow \frac{7}{16}$
                    \item pos $\rightarrow \frac{1}{2}$ 
                    \item t $\rightarrow \frac{1}{64}$
\end{enumerate}
\vspace{2 mm}
\begin{enumerate}
    \item $w_\text{l}$=$\frac{1}{16}$, $w_\text{r}$=$\frac{1}{16}$
    \begin{figure}[h]
            \centering
            {\includegraphics[width=6 cm]{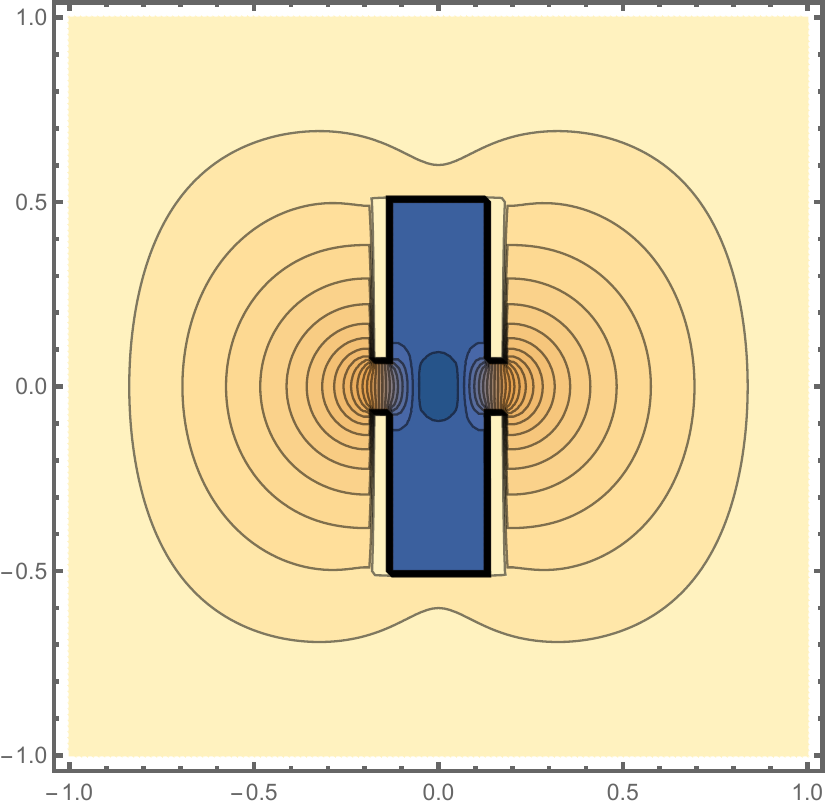} }
            \qquad
            {\includegraphics[width=6 cm]{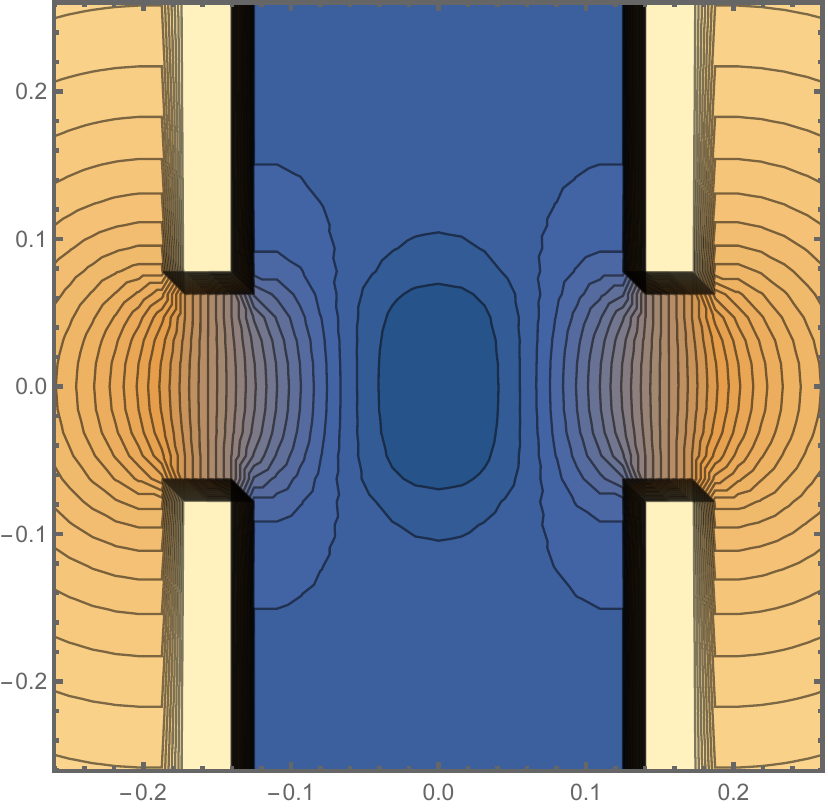}}
            \qquad
            \caption{P1}
            \label{fig:P1}
    \end{figure}
    \item $w_\text{l}$=$\frac{1}{16}$,$w_\text{r}$=$\frac{5}{16}$
    \begin{figure}[h]
            \centering
            {\includegraphics[width=6 cm]{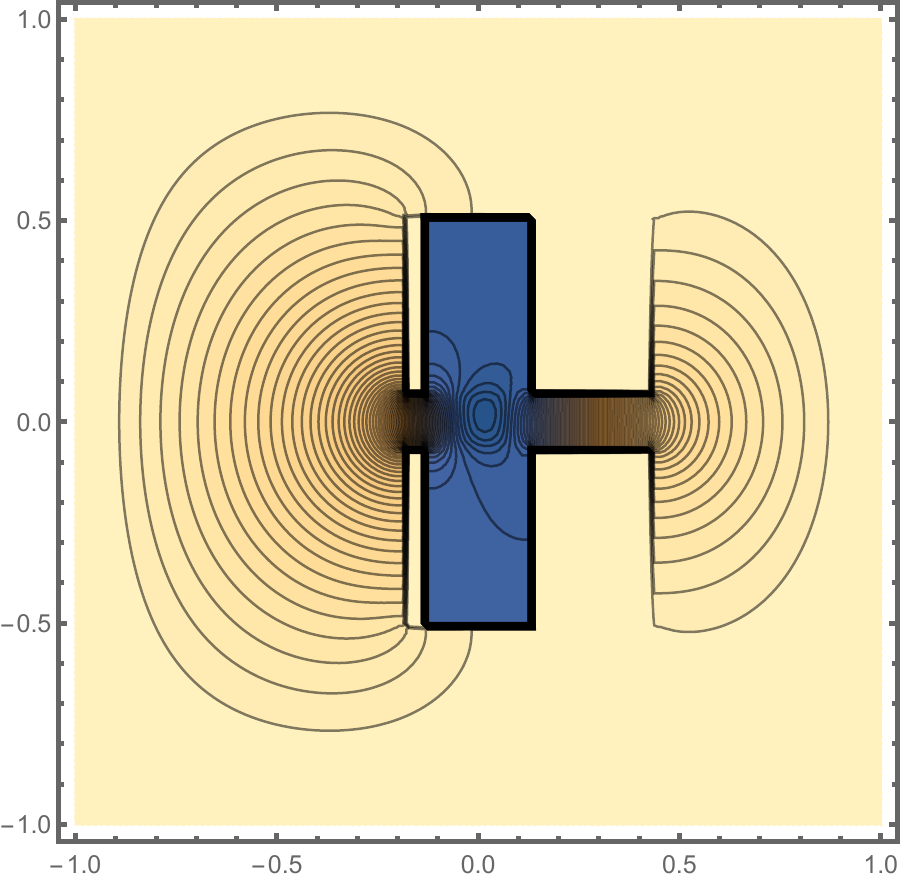} }
            \qquad
            {\includegraphics[width=6 cm]{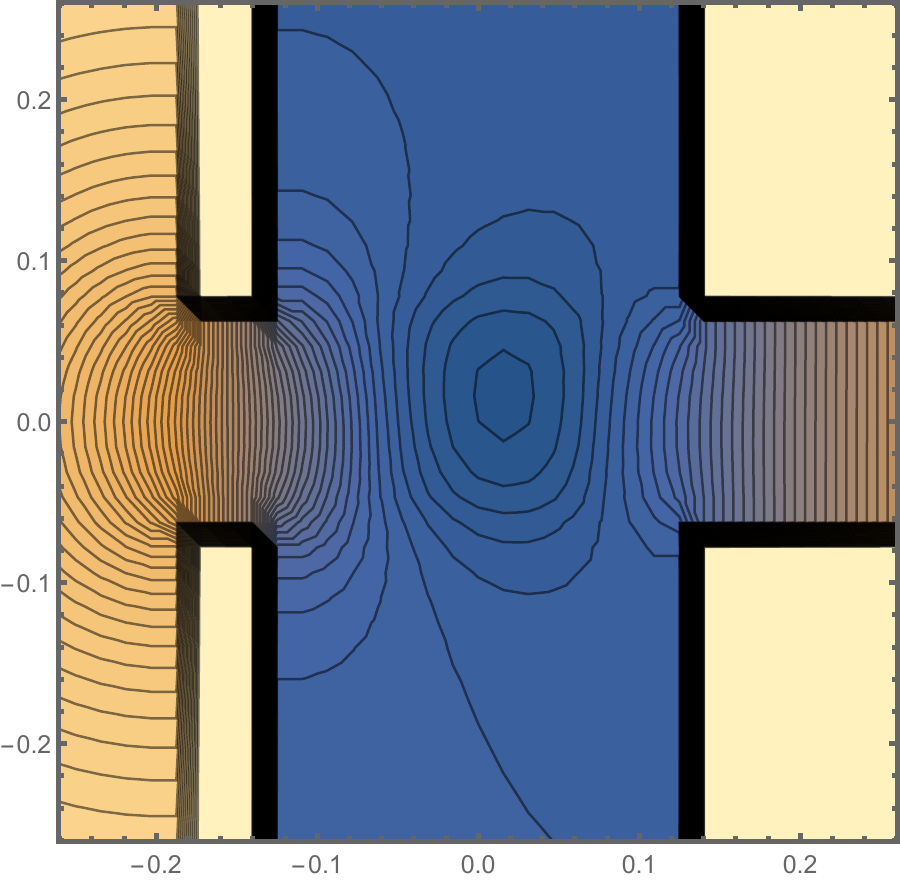}}
            \qquad
            \caption{P2}
            \label{fig:P2}
    \end{figure}
    \newpage
    \item $w_\text{l}$=$\frac{3}{16}$, $w_\text{r}$=$\frac{3}{16}$
    \begin{figure}[h]
            \centering
            {\includegraphics[width=6 cm]{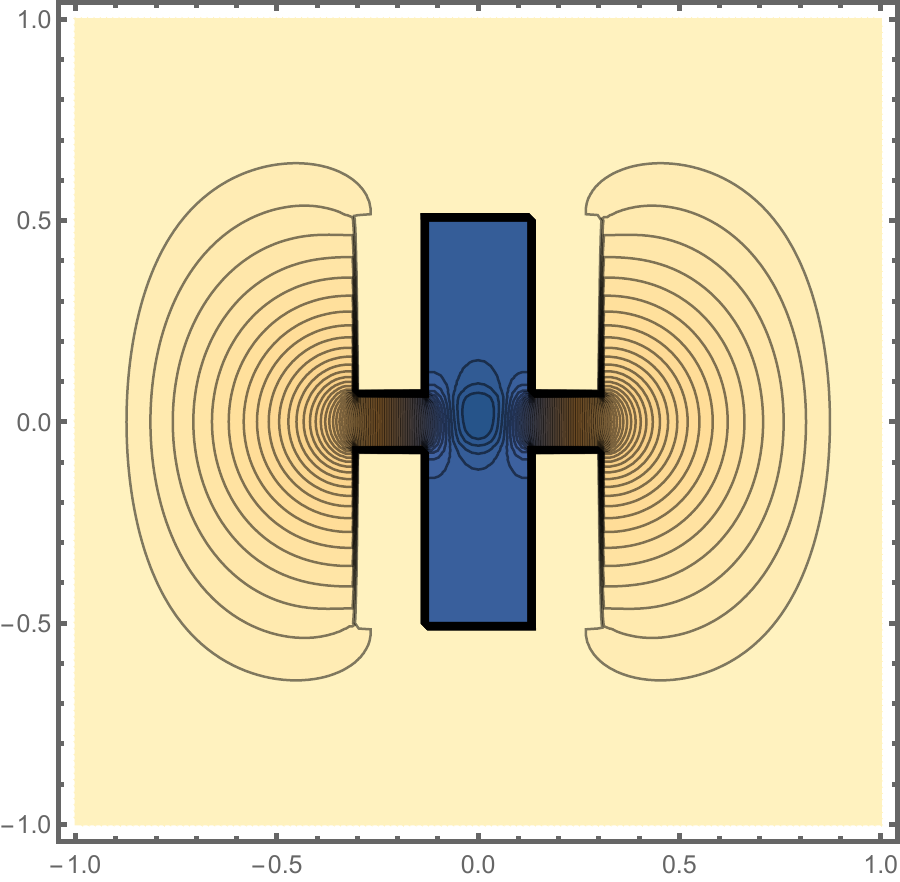} }
            \qquad
            {\includegraphics[width=6 cm]{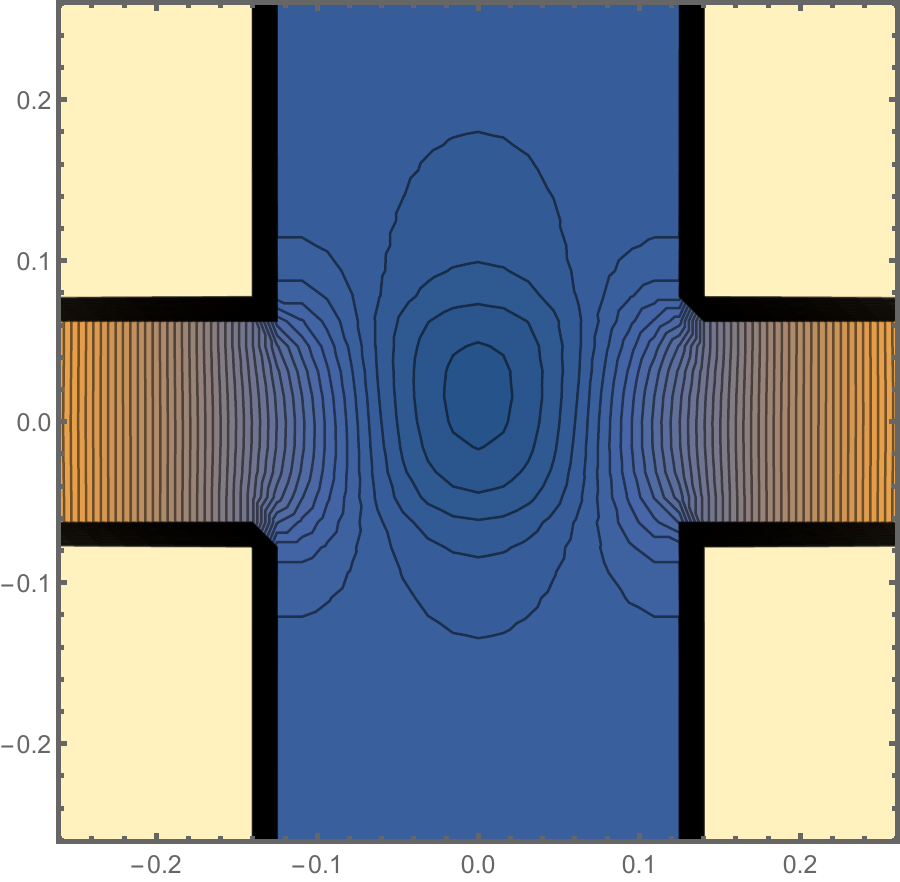}}
            \qquad
            \caption{P3}
            \label{fig:P3}
    \end{figure}
    \item $w_\text{l}$=$\frac{3}{16}$, $w_\text{r}$=$\frac{5}{16}$
    \begin{figure}[h]
            \centering
            {\includegraphics[width=6 cm]{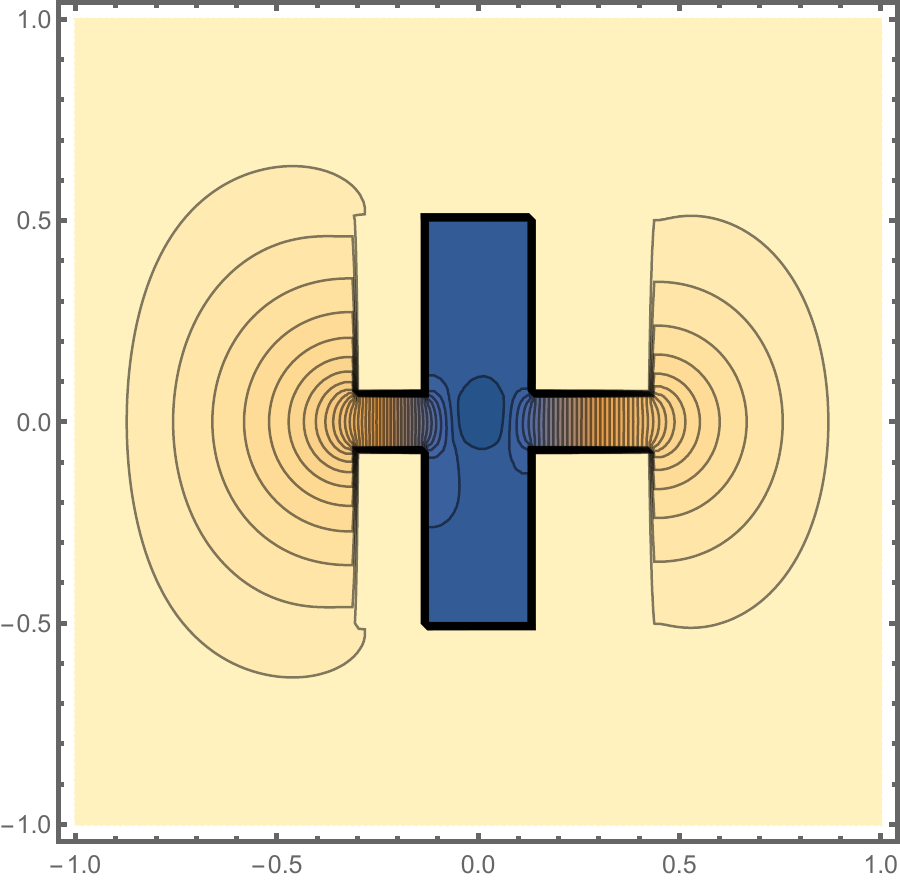} }
            \qquad
            {\includegraphics[width=6 cm]{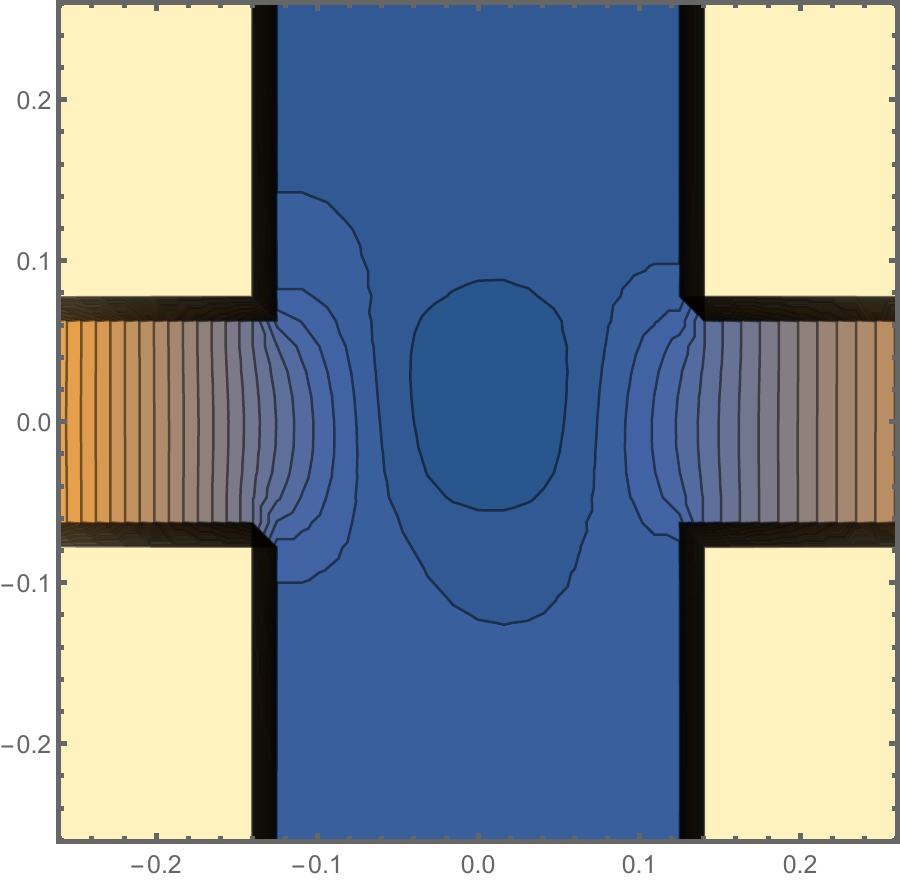}}
            \qquad
            \caption{P4}
            \label{fig:P4}
    \end{figure}
\end{enumerate}
\vspace{2 mm}
\subsection{Contour plots of the stream function contrary to the predictions of You et al.}
\label{sec:15.3}
\vspace{2mm}
\begin{itemize}
                    \item lsc $\rightarrow \frac{1}{8}$
                    \item pos $\rightarrow \frac{1}{4}$ - The y-coordinate that specifies the position locale of the superconductor
                    \item t $\rightarrow \frac{1}{64}$
                    \item d $\rightarrow \frac{14}{64}$ - so that the gap between the upper and the lower superconductors will be $\frac{1}{32}$
                    \item $w_\text{l}$ = $\frac{3}{8}$
                    \item $w_\text{r}$ = $\frac{3}{8}$
\end{itemize}
As expected, $\frac{C_2}{C_1} = 1$.
\newpage
\begin{enumerate}
    \item $\{B_\text{xextentn},B_\text{xextentp}\} \otimes \{B_\text{yextentn},B_\text{yextentp}\}\rightarrow \{\frac{6}{64},\frac{7}{64}\} \otimes \{\frac{-1}{16},\frac{1}{16}\}$
    \begin{figure}[h]
            \centering
            {\includegraphics[width=6 cm]{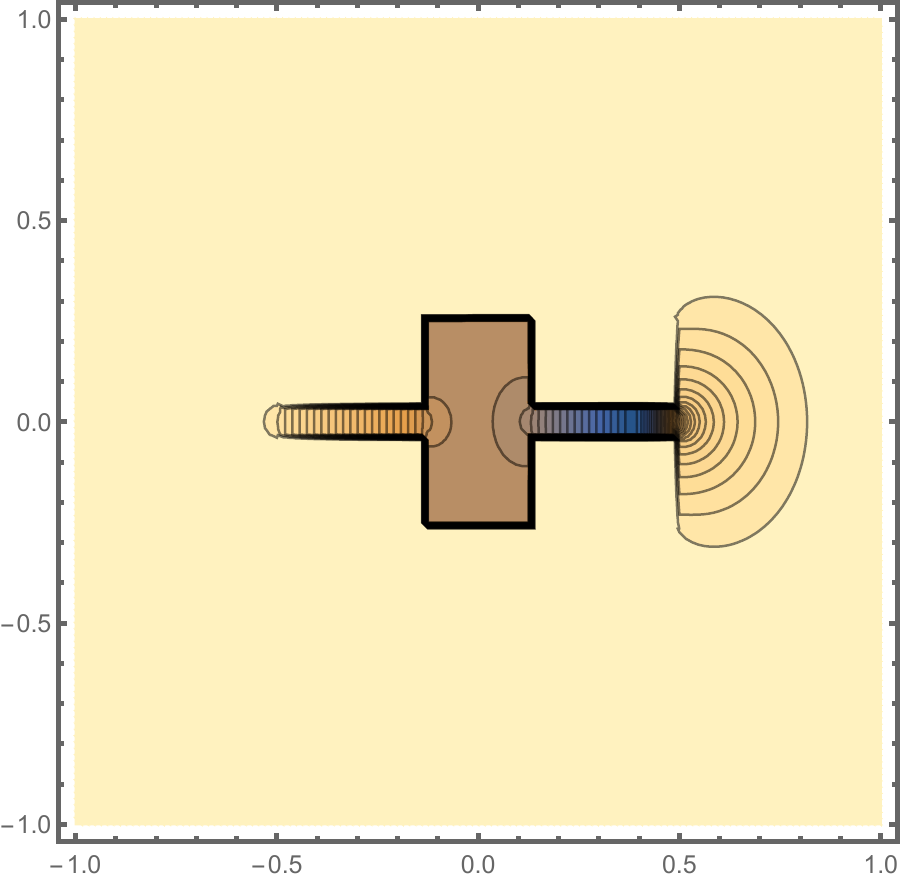} }
            \qquad
            {\includegraphics[width=6 cm]{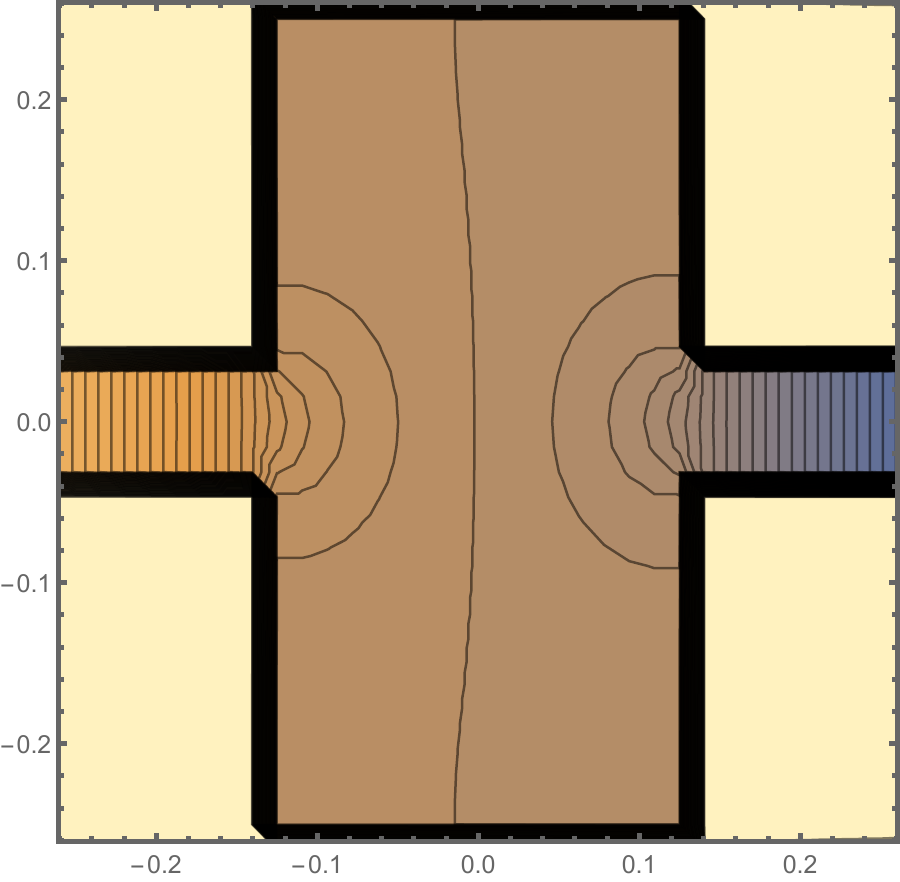}}
            \qquad
            \caption{P5}
            \label{fig:P5}
    \end{figure}
    \item $\{B_\text{xextentn},B_\text{xextentp}\}\otimes\{B_\text{yextentn},B_\text{yextentp}\}\rightarrow \{\frac{6}{64},\frac{7}{64}\} \otimes \{0,\frac{1}{16}\}$
    \begin{figure}[h]
            \centering
            {\includegraphics[width=6 cm]{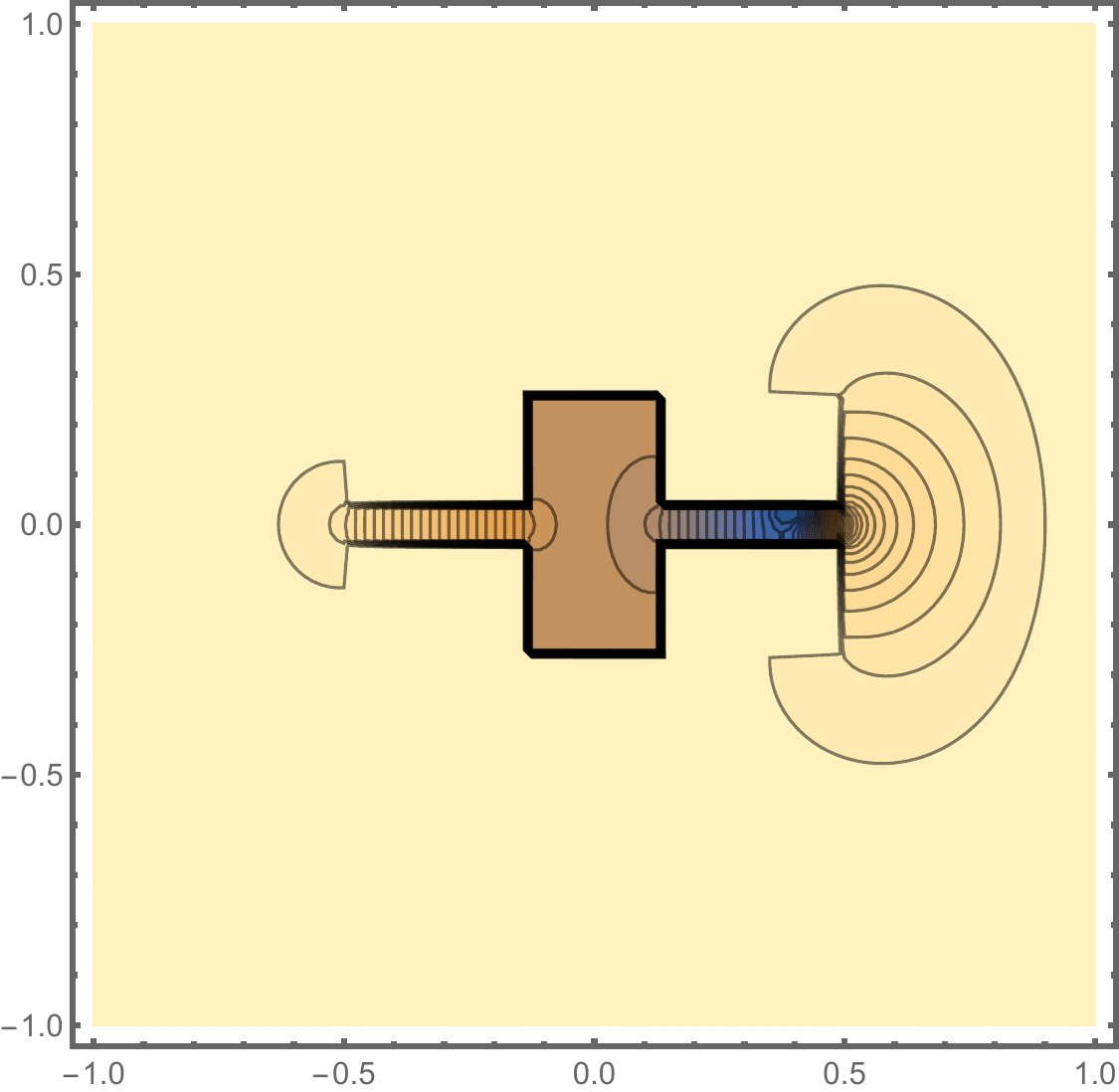} }
            \qquad
            {\includegraphics[width=6 cm]{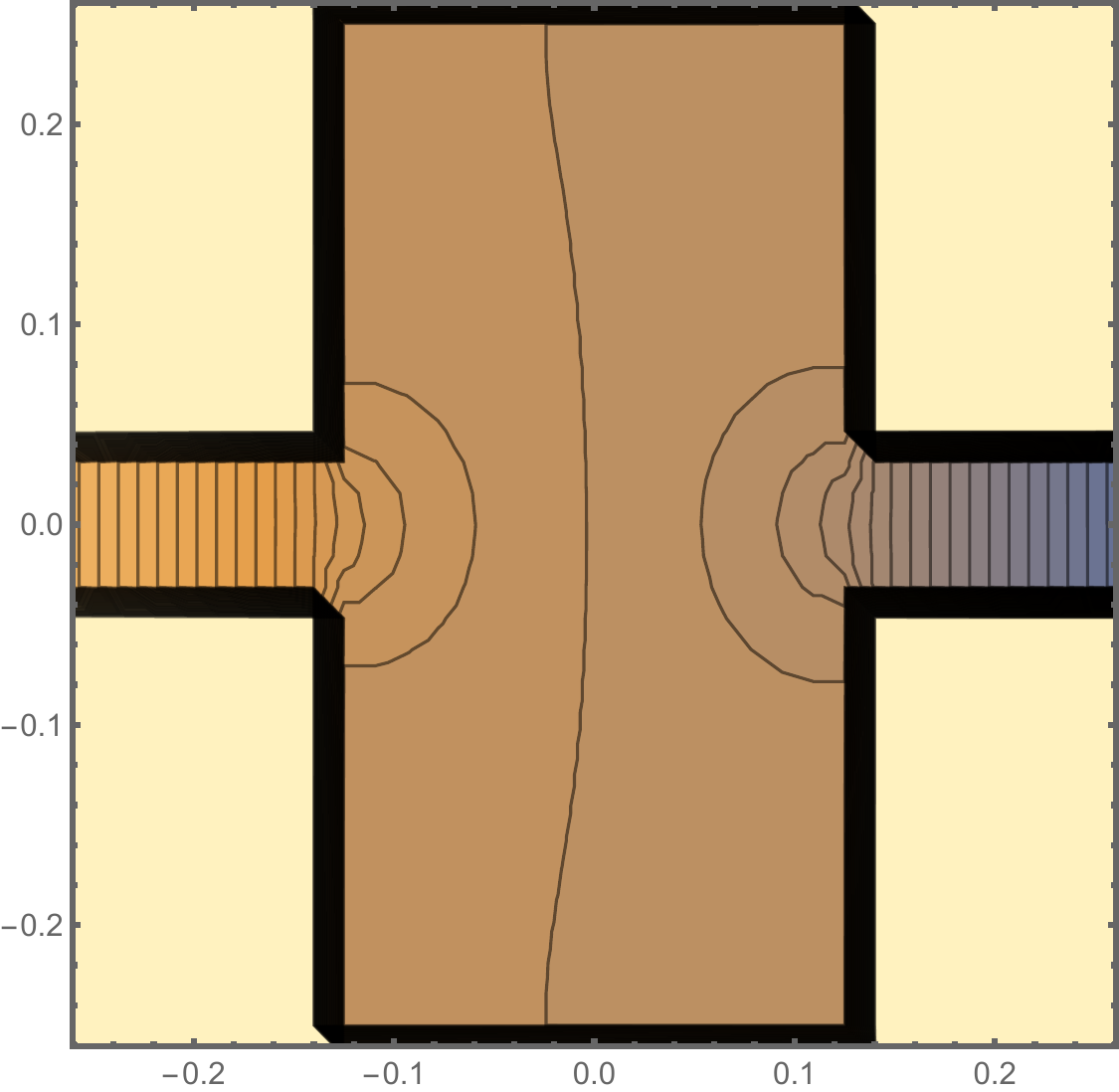}}
            \qquad
            \caption{P6}
            \label{fig:P6}
    \end{figure}
    \item $\{B_\text{xextentn},B_\text{xextentp}\}\otimes\{B_\text{yextentn},B_\text{yextentp}\}\rightarrow \{\frac{2}{8},\frac{3}{8}\} \otimes \{\frac{-1}{64},\frac{1}{64}\}$
    \begin{figure}[h]
            \centering
            {\includegraphics[width=6 cm]{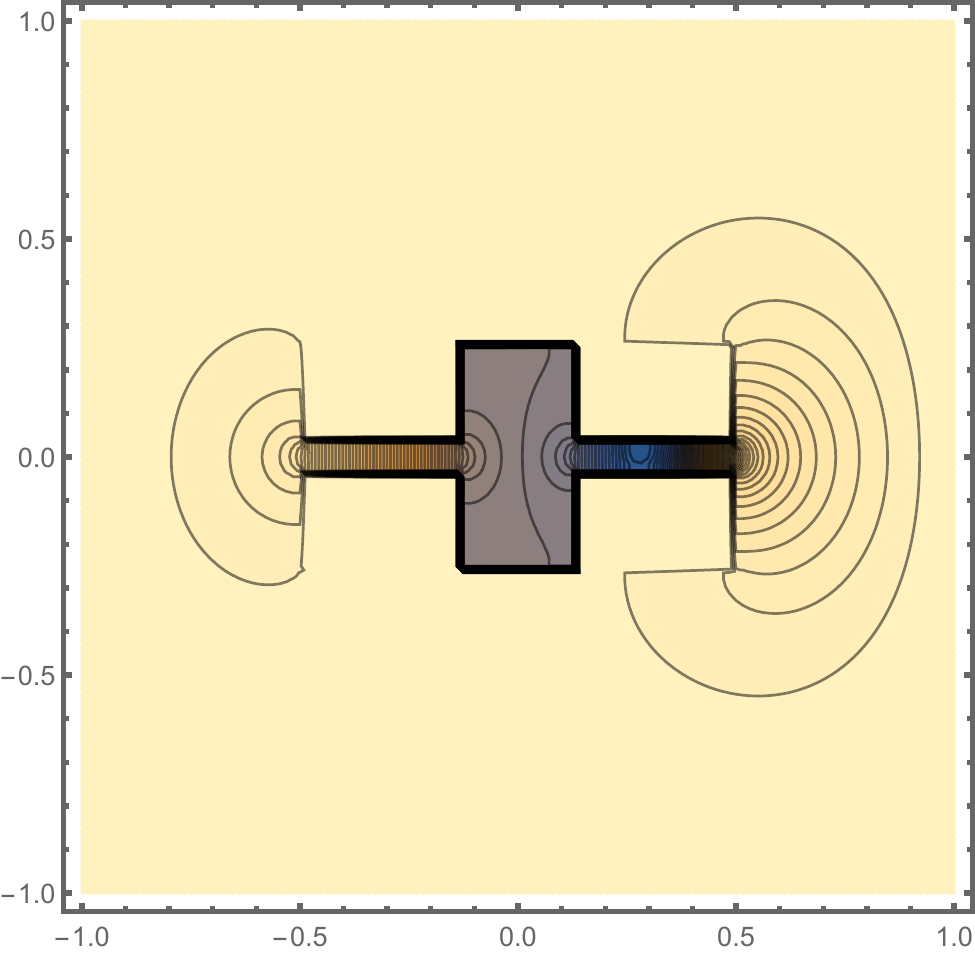} }
            \qquad
            {\includegraphics[width=6 cm]{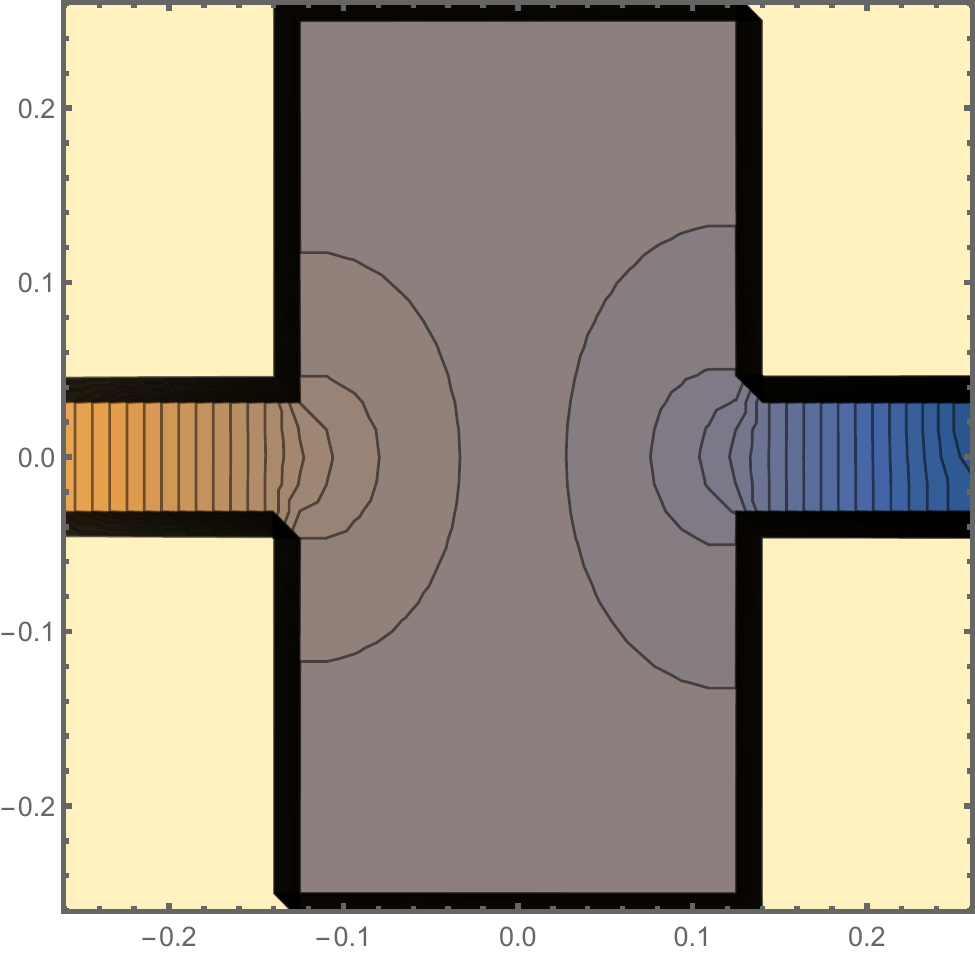}}
            \qquad
            \caption{P7}
            \label{fig:P7}
    \end{figure}
    \newpage
    \item $\{B_\text{xextentn},B_\text{xextentp}\}\otimes\{B_\text{yextentn},B_\text{yextentp}\}\rightarrow \{\frac{14}{64},\frac{22}{64}\} \otimes \{\frac{-1}{64},\frac{1}{64}\}$
    \begin{figure}[h]
            \centering
            {\includegraphics[width=6 cm]{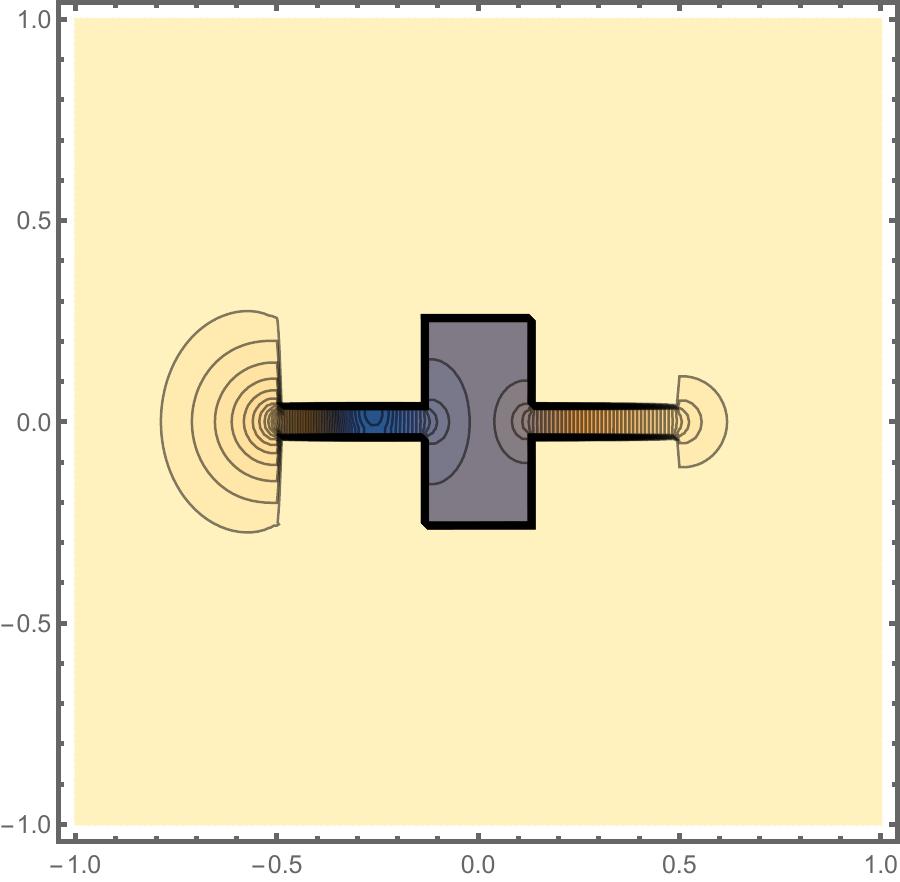} }
            \qquad
            {\includegraphics[width=6 cm]{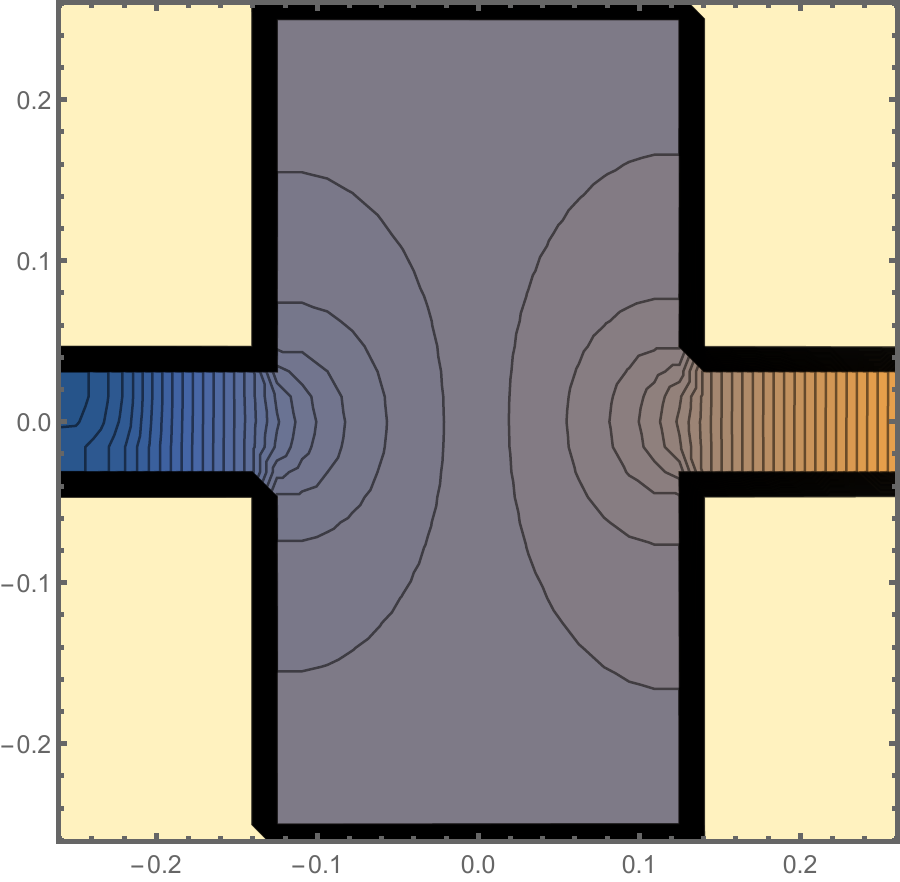}}
            \qquad
            \caption{P8}
            \label{fig:P8}
    \end{figure}
\vspace{2mm}
\\
From the contour plots in Fig.\hyperref[fig:P7]{P7} and Fig.\hyperref[fig:P8]{P8} as contrasted with the plots in Fig.\hyperref[fig:P5]{P5} and Fig.\hyperref[fig:P6]{P6}, it is evident that only a minimal deviation from the contour plot as predicted by Ref.~\hyperref[sec:Ref]{[2]} viz. Fig.\hyperref[fig:P3]{P3}, is seen as long as the magnetic field is confined to the U-shaped region. On placing the field in the gap between the two plates, clearly a large deviation results.
\vspace{2mm}
\\
However, as mentioned before, note that the fact that this deviation from the predictions of Ref.~\hyperref[sec:Ref]{[2]} is comparitively less is \textbf{not} due to the accuracy of the prediction itself but due to the physical structure of the continuous geometry in which the U-shaped region is small. This statement is supported by evidence in the next section.
\subsection{Contrary to the predictions of You et al.: larger deviations}
\label{sec:15.4}
We stated in Sec.~\hyperref[sec:13.2.1]{13.2.1} that the relatively small deviation of $\approx$ 0.1 from the predictions of You et al. when the field was placed in the `U-shaped' region is due to the small dimensions of the structure. 
\vspace{2mm}
We validate that statement by tabulating the results of a set of simulations on a larger structure :\textemdash
\vspace{1mm}
\begin{enumerate}
                    \item (xmin,xmax) $\rightarrow$ (-2,2)
                    \item (ymin,ymax) $\rightarrow$ (-2,2)
                    \item lsc $\rightarrow$ 1
                    \item pos $\rightarrow \frac{1}{4}$ 
                    \item t $\rightarrow \frac{1}{64}$
                    \item d $\rightarrow \frac{14}{64}$
                    \item $w_\text{l}$ = $\frac{3}{8}$
                    \item $w_\text{r}$ = $\frac{3}{8}$
\end{enumerate}
As expected, $\frac{C_2}{C_1} = 1$.
\vspace{2mm}

                \begin{tabular}{||c c c c c||}
                  \hline
                  $B_\text{xextentn}$ & $B_\text{xextentp}$ & $B_\text{yextentn}$ & $B_\text{yextentp}$ & $\frac{\alpha_1}{\alpha_2}$ IM \\ [1.0  ex]     
                  
                  \hline \hline
                  $\frac{4}{8}$ & $\frac{7}{8}$ & $\frac{-1}{16}$ & $\frac{1}{16}$ & 0.815\\ [1 ex]
                   \hline
                  $\frac{4}{8}$ & $\frac{7}{8}$ & 0 & $\frac{1}{16}$ & 0.815\\ [1 ex]
                   \hline
                  $\frac{4}{8}$ & $\frac{7}{8}$ & $\frac{2}{16}$ & $\frac{3}{16}$ & 0.817\\ [1 ex]
                   \hline
                  $\frac{1}{8}$ & $\frac{7}{8}$ & $\frac{-3}{16}$ & $\frac{3}{16}$ & 0.863\\ [1 ex]
                  \hline
                  \hline
                \end{tabular}
\end{enumerate}
Thus, there is a deviation of the order of $2*10^{-1}$ in the values of $\frac{\alpha_1}{\alpha_2}$ from the expected value of 1 - $\frac{C_2}{C_1}$. This is a markedly large deviation and an order of magnitude larger than the expected errors due to the finitude of the approximation and any truncation errors.
\vspace{2mm}
\\
Hence, there is numerical evidence that the maximum deviation observed from the predictions of You et al., when the uniform magnetic field is placed within the `U-shaped' concavity of the structure, increases as the size of the concavity increases - thereby allowing us to create a greater `asymmetry' with the flux with respect to the overall symmetry of the structure.
\newpage
\subsection{Vector plots of the magnetic vector potential}
For each of the representative plots in Sec.~\hyperref[sec:15.1]{15.1} and Sec.~\hyperref[sec:15.2]{15.2}, we plot the corresponding magnetic vector potentials.
\vspace{2mm}
\subsubsection{In accordance with the predictions of You et al.}

\begin{enumerate}
                    \item $\{ \text{B}_{xextentn},\text{B}_{xextentp} \}$ $\rightarrow$ $\{ \frac{-1}{16},\frac{1}{16}\}$
                    \item $\{ \text{B}_{yextentn},\text{B}_{yextentp} \}$ $\rightarrow$ $\{ \frac{-1}{16},\frac{1}{16}\}$
                    \item lsc $\rightarrow \frac{1}{8}$
                    \item d $\rightarrow \frac{7}{16}$
                    \item pos $\rightarrow \frac{1}{2}$ 
                    \item t $\rightarrow \frac{1}{64}$
\end{enumerate}
\vspace{2mm}
\begin{enumerate}
    \item $w_\text{l}$=$\frac{1}{16}$, $w_\text{r}$=$\frac{1}{16}$
    \begin{figure}[h]
            \centering
            {\includegraphics[width=6 cm]{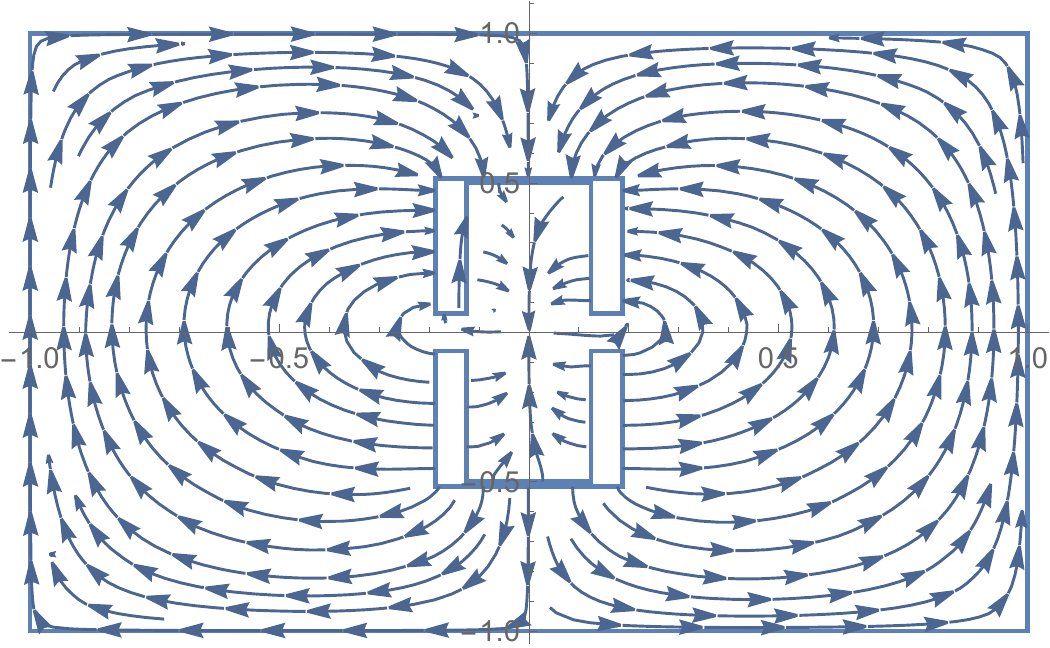} }
            \qquad
            {\includegraphics[width=6 cm]{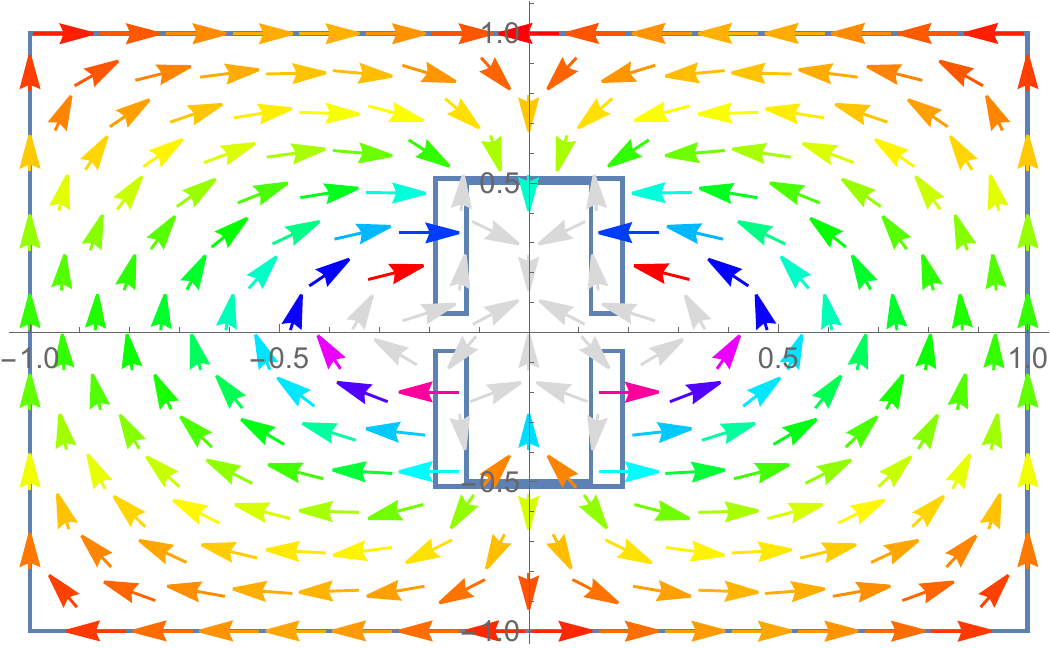}}
            \qquad
            \caption{P9}
            \label{fig:P9}
    \end{figure}
    \item $w_\text{l}$=$\frac{1}{16}$,$w_\text{r}$=$\frac{5}{16}$
    \begin{figure}[h]
            \centering
            {\includegraphics[width=6 cm]{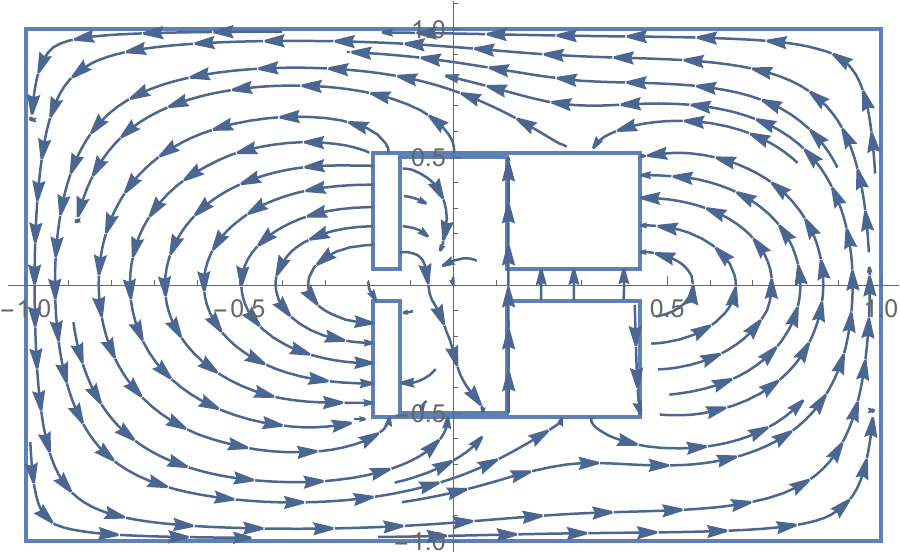} }
            \qquad
            {\includegraphics[width=6 cm]{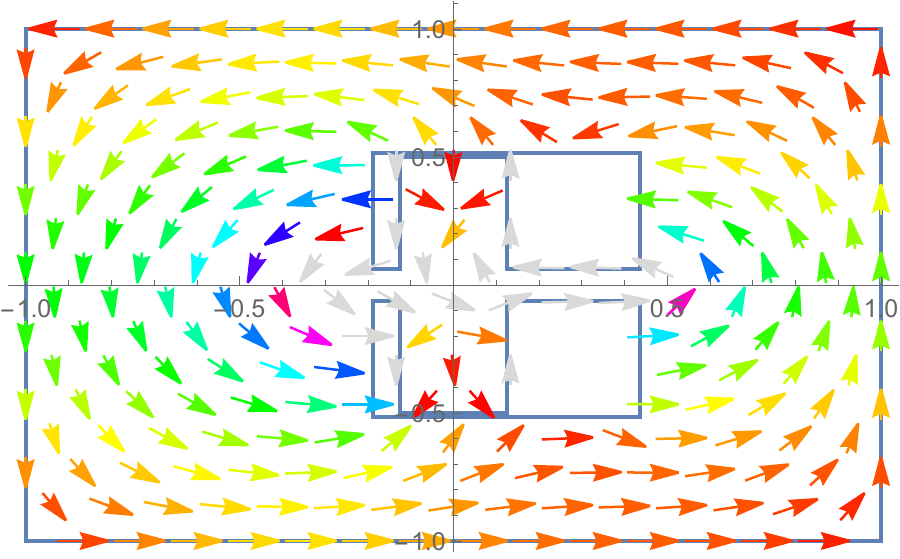}}
            \qquad
            \caption{P10}
            \label{fig:P10}
    \end{figure}
    \item $w_\text{l}$=$\frac{3}{16}$, $w_\text{r}$=$\frac{3}{16}$
    \begin{figure}[h]
            \centering
            {\includegraphics[width=6 cm]{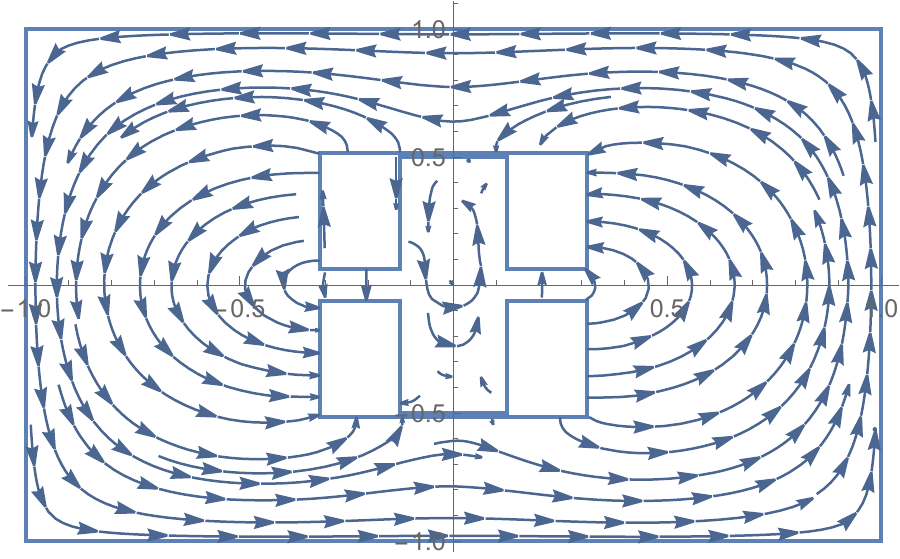} }
            \qquad
            {\includegraphics[width=6 cm]{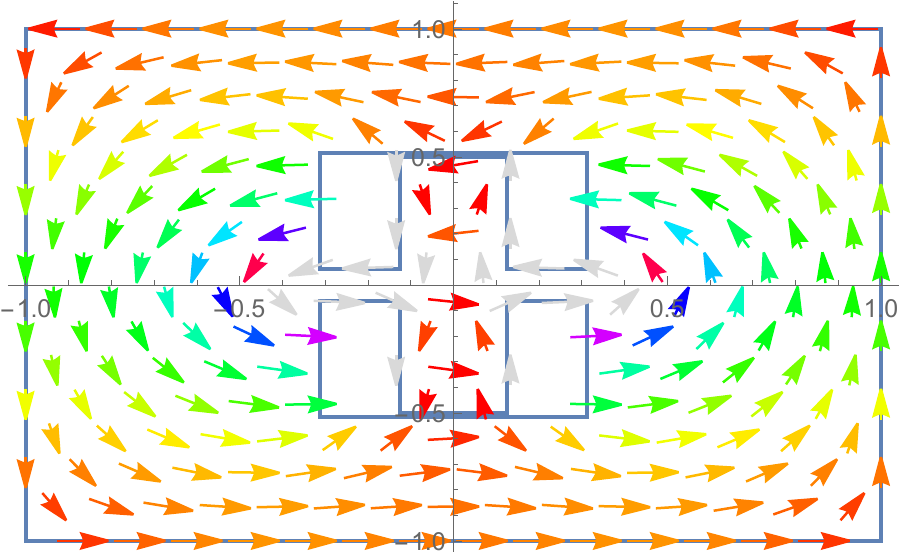}}
            \qquad
            \caption{P11}
            \label{fig:P11}
    \end{figure}
    \newpage
    \item $w_\text{l}$=$\frac{3}{16}$, $w_\text{r}$=$\frac{5}{16}$
    \begin{figure}[h]
            \centering
            {\includegraphics[width=6 cm]{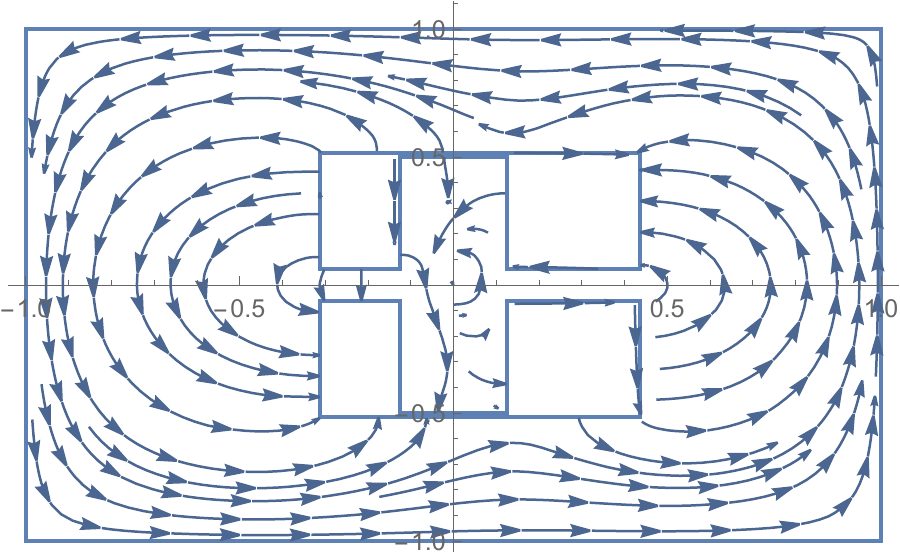} }
            \qquad
            {\includegraphics[width=6 cm]{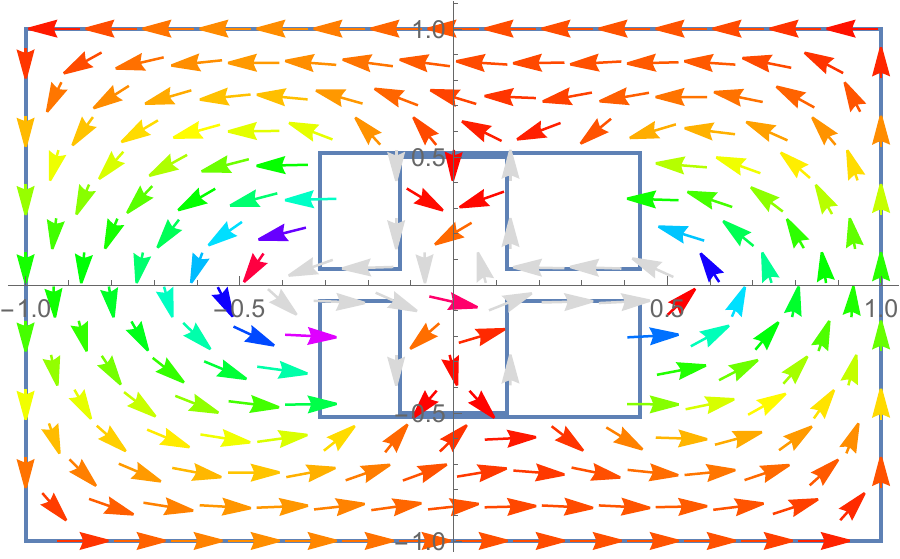}}
            \qquad
            \caption{P12}
            \label{fig:P12}
    \end{figure}
\end{enumerate}
\vspace{1mm}
\subsubsection{Contrary to the predictions of You et al.}
\begin{enumerate}
                    \item lsc $\rightarrow \frac{1}{8}$
                    \item pos $\rightarrow \frac{1}{4}$ - The y-coordinate that specifies the position locale of the superconductor
                    \item t $\rightarrow \frac{1}{64}$
                    \item d $\rightarrow \frac{14}{64}$ - so that the gap between the upper and the lower superconductors will be $\frac{1}{32}$
                    \item $w_\text{l}$ = $\frac{3}{8}$
                    \item $w_\text{r}$ = $\frac{3}{8}$
\end{enumerate}
$\therefore \frac{C_2}{C_1} = 1$
\vspace{2 mm}
\begin{enumerate}
    \item $\{B_\text{xextentn},B_\text{xextentp}\}\otimes\{B_\text{yextentn},B_\text{yextentp}\}\rightarrow \{\frac{6}{64},\frac{7}{64}\} \otimes \{\frac{-1}{16},\frac{1}{16}\}$
    \begin{figure}[h]
            \centering
            {\includegraphics[width=6 cm]{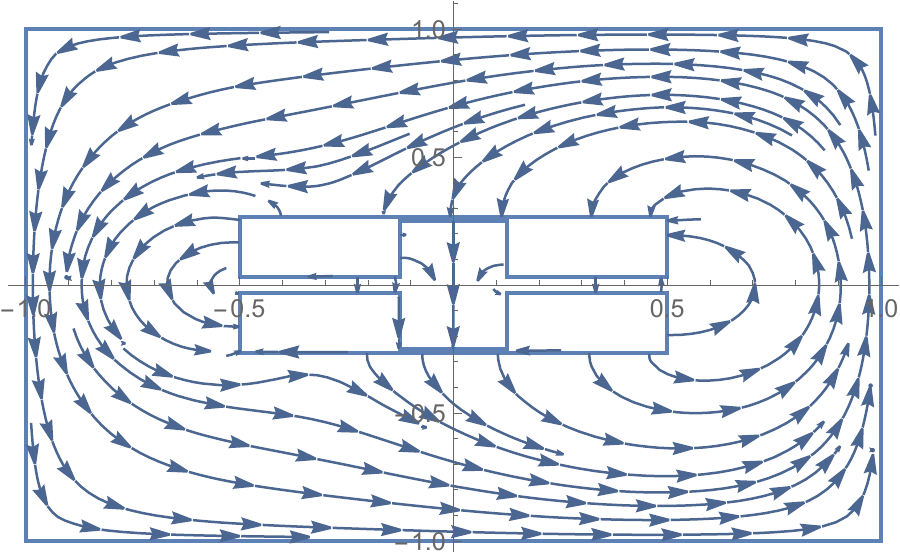} }
            \qquad
            {\includegraphics[width=6 cm]{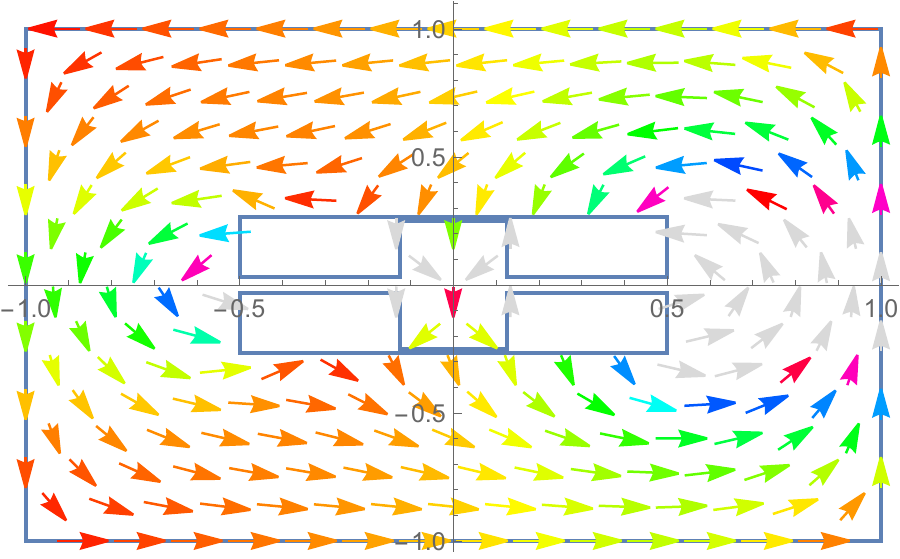}}
            \qquad
            \caption{P13}
            \label{fig:P13}
    \end{figure}
    \item $\{B_\text{xextentn},B_\text{xextentp}\}\otimes\{B_\text{yextentn},B_\text{yextentp}\}\rightarrow \{\frac{6}{64},\frac{7}{64}\} \otimes \{0,\frac{1}{16}\}$
    \begin{figure}[h]
            \centering
            {\includegraphics[width=6 cm]{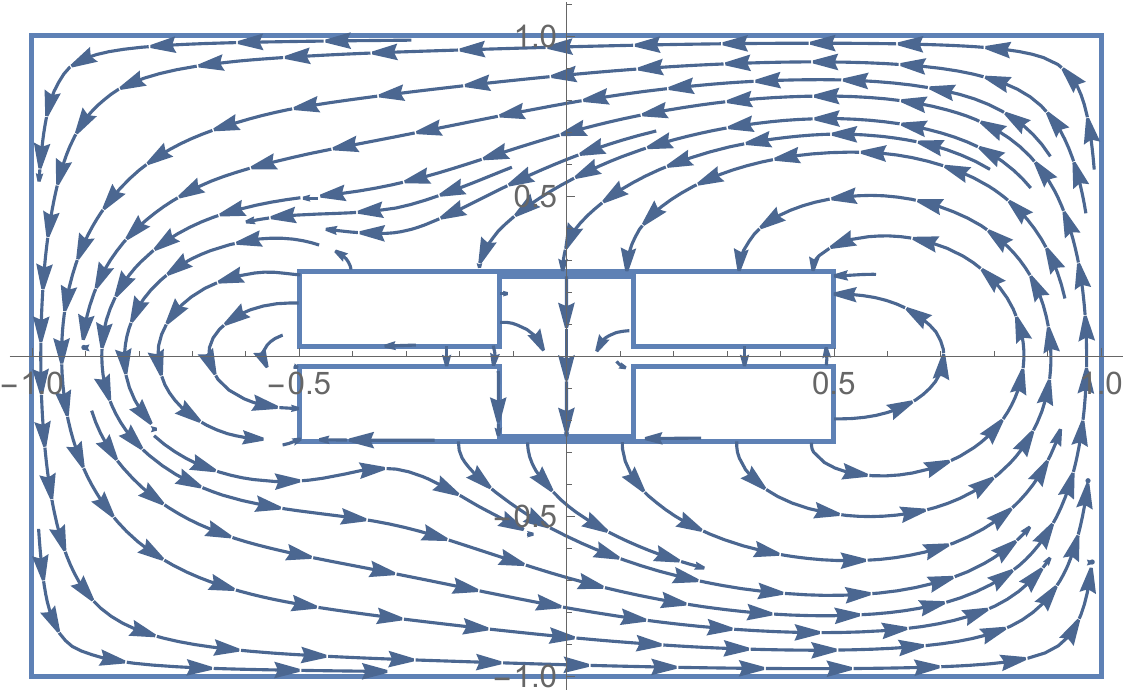} }
            \qquad
            {\includegraphics[width=6 cm]{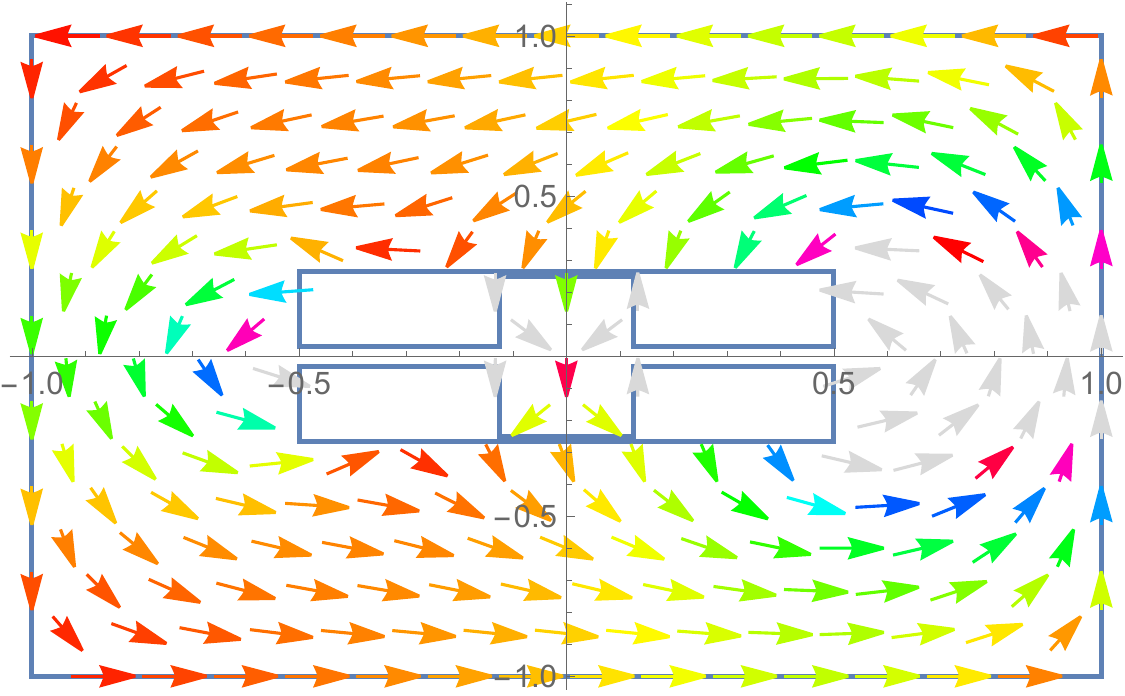}}
            \qquad
            \caption{P14}
            \label{fig:P14}
    \end{figure}
    \newpage
    \item $\{B_\text{xextentn},B_\text{xextentp}\}\otimes\{B_\text{yextentn},B_\text{yextentp}\}\rightarrow \{\frac{2}{8},\frac{3}{8}\} \otimes \{\frac{-1}{64},\frac{1}{64}\}$
    \begin{figure}[h]
            \centering
            {\includegraphics[width=6 cm]{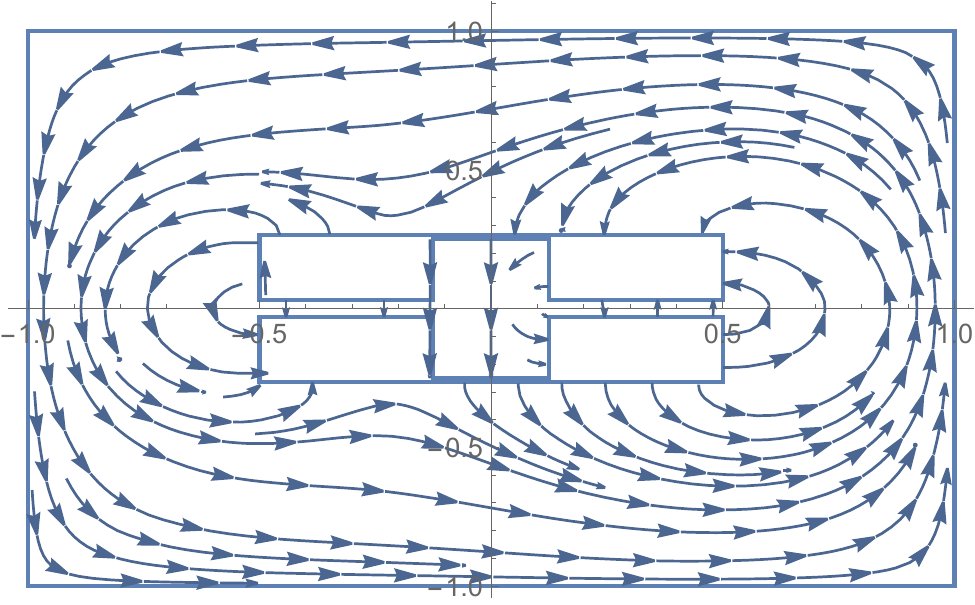} }
            \qquad
            {\includegraphics[width=6 cm]{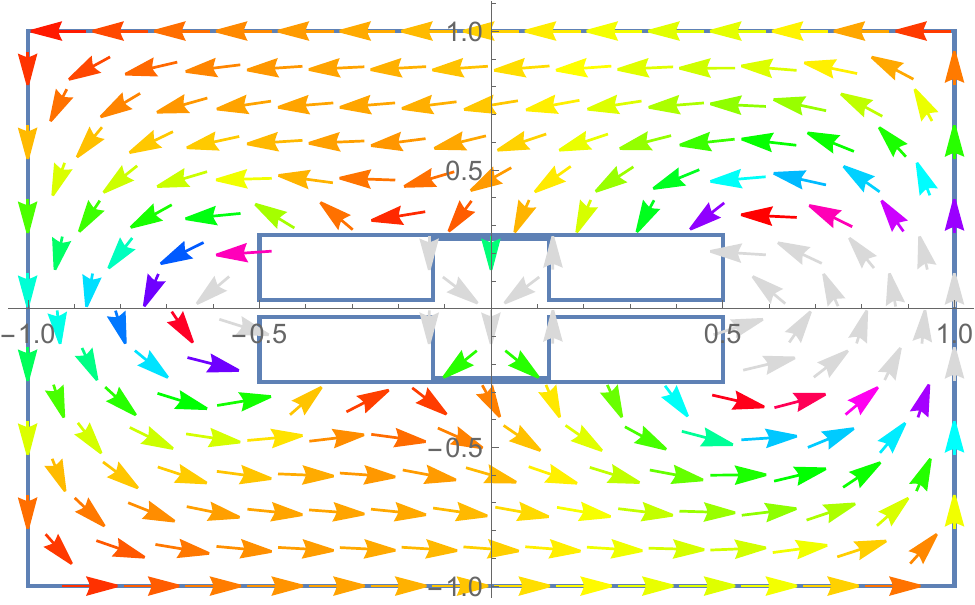}}
            \qquad
            \caption{P15}
            \label{fig:P15}

\flushleft  \hspace{4 mm}4. $\{B_\text{xextentn},B_\text{xextentp}\}\otimes\{B_\text{yextentn},B_\text{yextentp}\}\rightarrow \{\frac{14}{64},\frac{22}{64}\} \otimes \{\frac{-1}{64},\frac{1}{64}\}$
    \newline
    \\
            \centering
            {\includegraphics[width=6 cm]{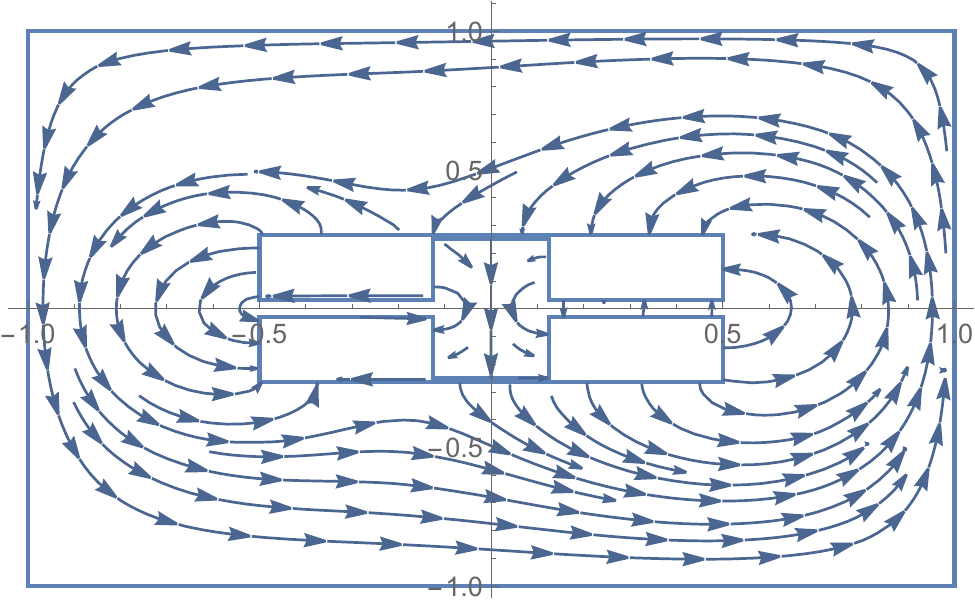} }
            \qquad
            {\includegraphics[width=6 cm]{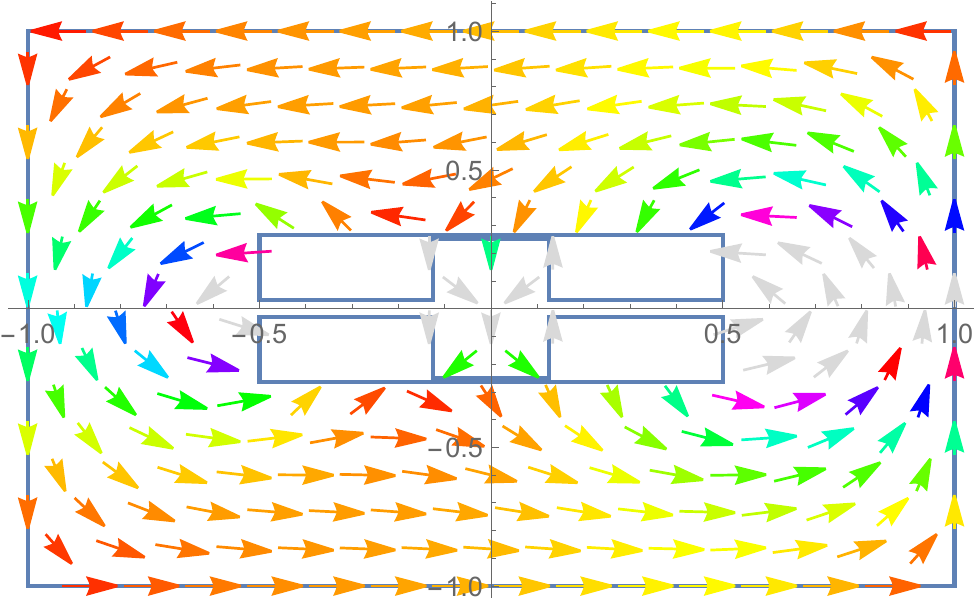}}
            \qquad
            \caption{P16}
            \label{fig:P16}
    \end{figure}
\end{enumerate}
\vspace{4mm}
\subsubsection{Comparative analysis of electrostatics and magnetostatics via their vector plots}
The calculation of the electrostatic junction capacitances is performed by the standard method of assigning two distinct, constant potentials to the island and the ground and after a finite difference analysis to calculate the electrostatic potentials, the ratio of the capacitances is derived as a ratio of the corresponding Gaussian surface integrals. 
\vspace{2mm}
Thus the finite difference analysis performed to evaluate the ratio of the capacitances gives us an insight into the nature of the electrostatics of the structure in question.
\vspace{2mm}
In this section, we contrast a set of the electrostatic plots by juxtaposing them alongside the corresponding magnetostatics plot. We observe that the distinctions between the two are quite evident - as expected from the behaviour of electric fields and the magnetic vector potentials.
\vspace{2 mm}
\begin{enumerate}
                    \item lsc $\rightarrow \frac{1}{8}$
                    \item pos $\rightarrow \frac{1}{4}$ - The y-coordinate that specifies the position locale of the superconductor
                    \item t $\rightarrow \frac{1}{64}$
                    \item d $\rightarrow \frac{14}{64}$ - so that the gap between the upper and the lower superconductors will be $\frac{1}{32}$
                    \item $w_\text{l}$ = $\frac{3}{8}$
                    \item $w_\text{r}$ = $\frac{3}{8}$
\end{enumerate}
\newpage
\begin{enumerate}
    \item The electrostatics plots
    \begin{figure}[h]
            \centering
            {\includegraphics[width=6 cm]{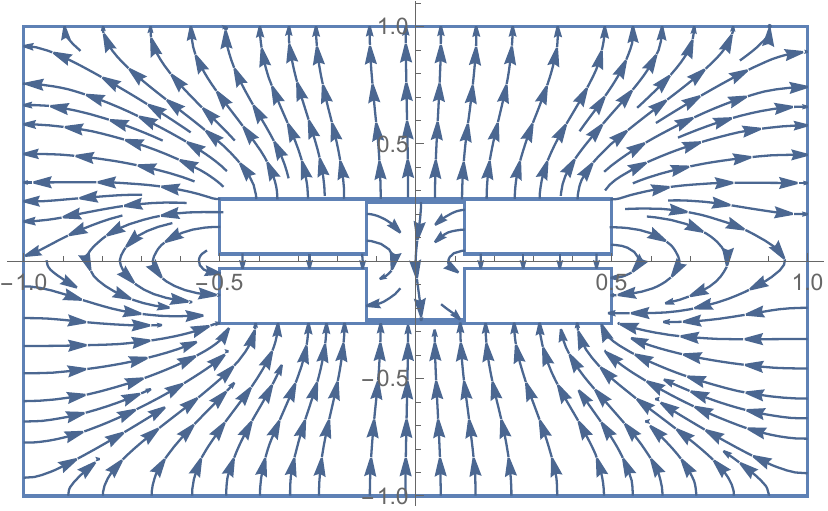} }
            \qquad
            {\includegraphics[width=6 cm]{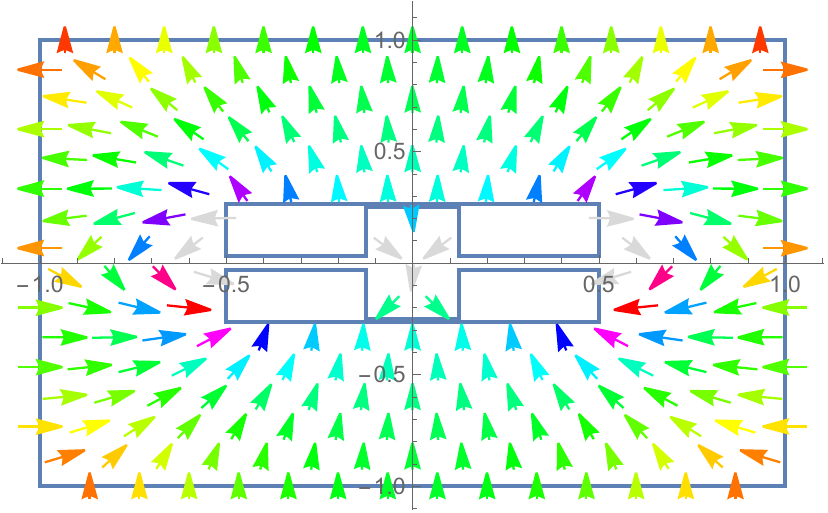}}
            \qquad
            \caption{P17}
            \label{fig:P17}
    \end{figure}
    \item The magnetostatics plots
    \begin{enumerate}
    
    \item $\{B_\text{xextentn},B_\text{xextentp}\}\otimes\{B_\text{yextentn},B_\text{yextentp}\}\rightarrow \{\frac{2}{8},\frac{3}{8}\} \otimes \{\frac{-1}{64},\frac{1}{64}\}$
    \begin{figure}[h]
            \centering
            {\includegraphics[width=6 cm]{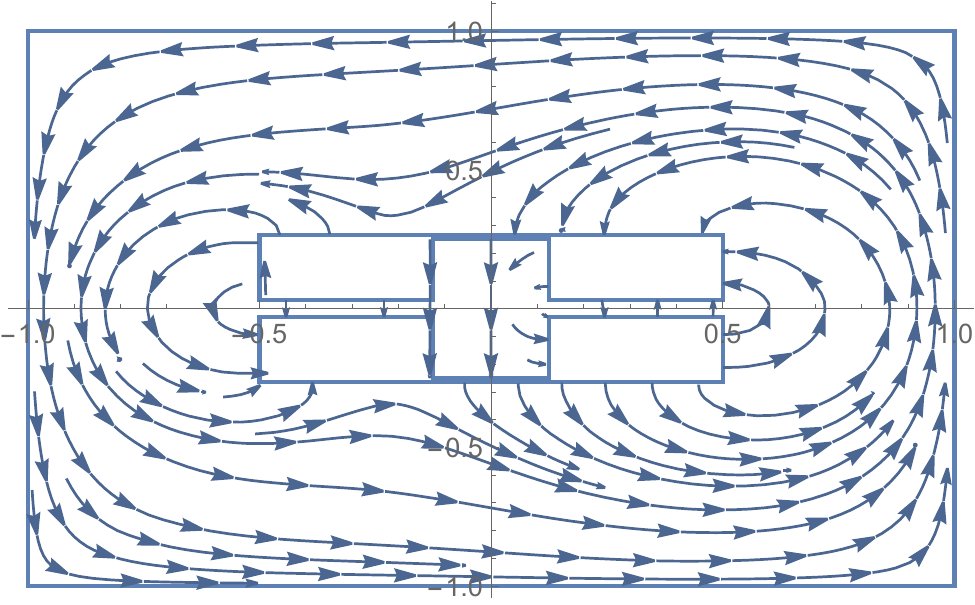} }
            \qquad
            {\includegraphics[width=6 cm]{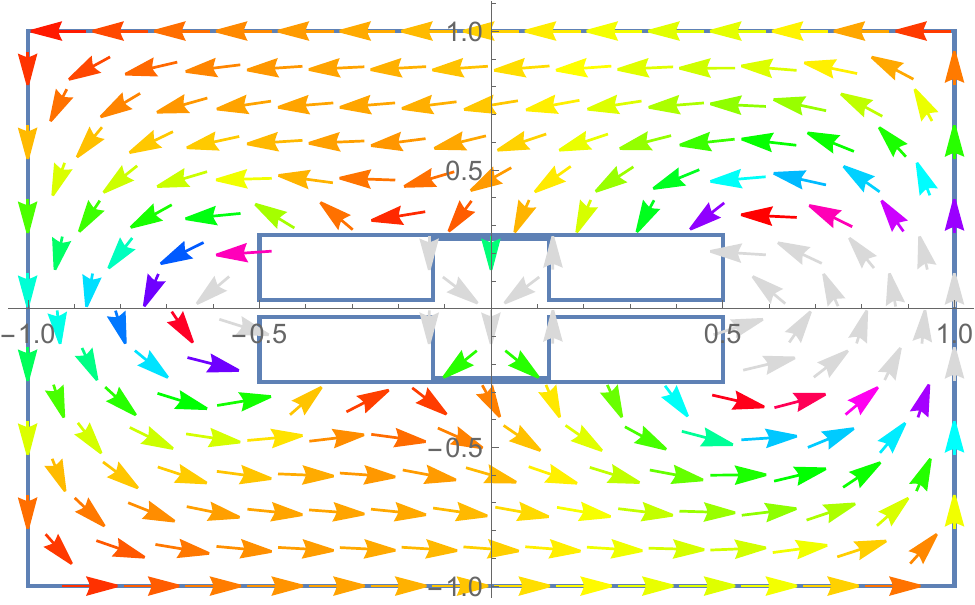}}
            \qquad
            \caption{P18}
            \label{fig:P18}
    \end{figure}
    \item $\{B_\text{xextentn},B_\text{xextentp}\}\otimes\{B_\text{yextentn},B_\text{yextentp}\}\rightarrow \{\frac{14}{64},\frac{22}{64}\} \otimes \{\frac{-1}{64},\frac{1}{64}\}$
    \begin{figure}[h]
            \centering
            {\includegraphics[width=6 cm]{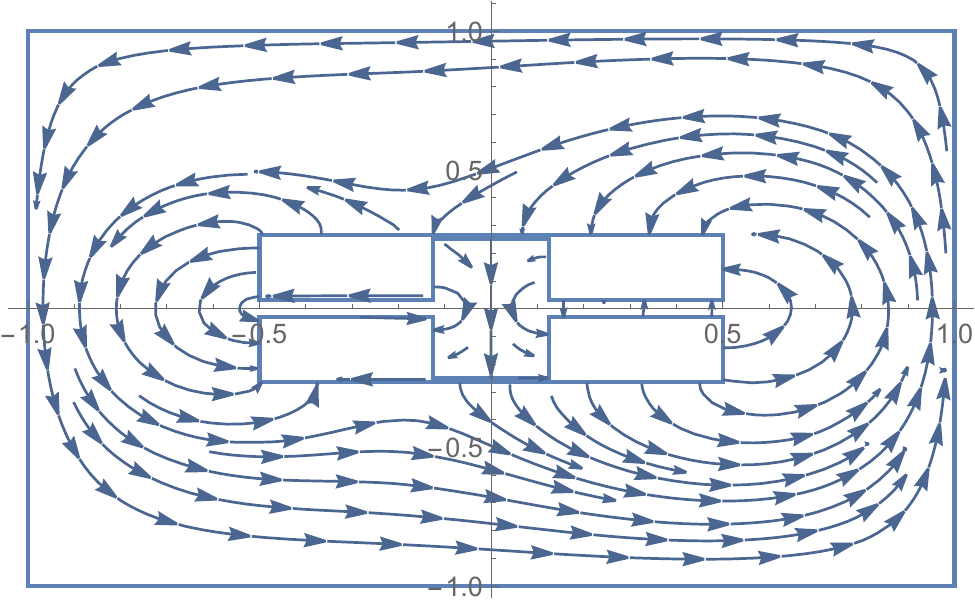} }
            \qquad
            {\includegraphics[width=6 cm]{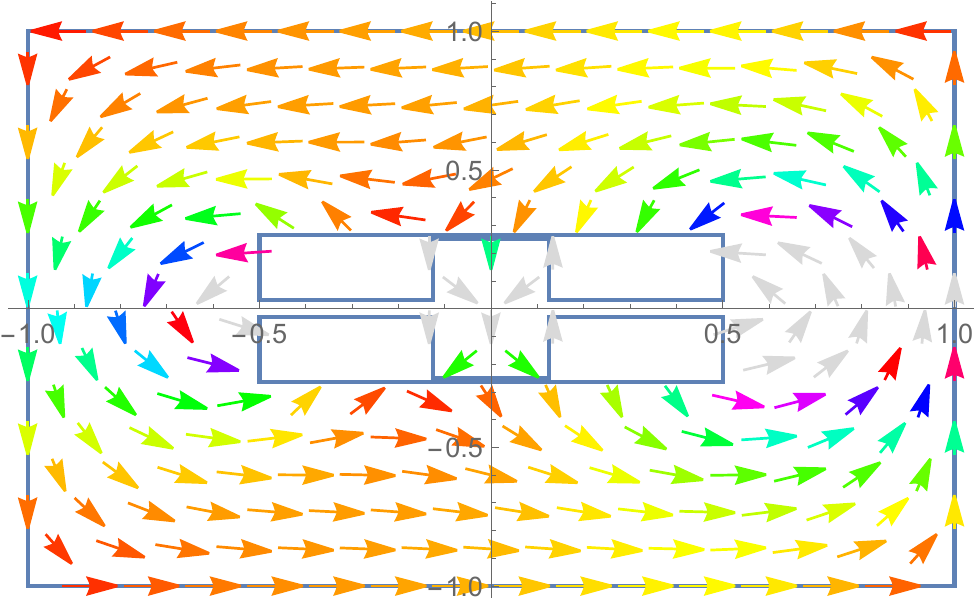}}
            \qquad
            \caption{P19}
            \label{fig:P19}
    \end{figure}
    \end{enumerate}
\end{enumerate}
\vspace{2 mm}
Analysis :-
\begin{enumerate}
    \item The magnetic vector potentials form closed loops at sufficiently large enough distances from the superconducting surfaces. 
    \\
    On the other hand, as expected, the electrostatics case does \textbf{not} have any closed loops and the field lines `emerge' from the positively charged island and terminate at the ground. Once again, those field lines at the boundary can be thought of as ones emerging from/terminating at the island/ground at infinity.
    \vspace{1mm}
    \item Close to the surfaces of the superconductors, \textbf{both} the magnetic vector potential and the electric field are perpendicular to the surface.
    \vspace{1mm}
    \item The magnetostatics plots depend on the locale of the magnetic flux as well as the structural configuration. However, the electrostatics plots depend solely on the structural configuration.
    \vspace{1mm}
    This is seen as maintaining structural symmetry upon reflection in the line x=0 ensures that the electrostatics plot bears the same symmetry while the magnetostatics plot, does not. 
    \vspace{1mm}
    \item The magnetic vector potential is tangential to the finite approximation of `$\infty$' whereas the electric field lines are perpendicular to the boundary.
\end{enumerate}
\hrule
\end{document}